\documentclass[%
 reprint,
showpacs,
 amsmath,amssymb,
 aps,floatfix
]{revtex4-1}

\usepackage{graphicx}
\usepackage{dcolumn}
\usepackage{bm}

\begin{document}


\title{ Dyakonov Waves in Biaxial Anisotropic Crystals}

\author{Evgenii E. Narimanov}
\affiliation{School of Electrical and Computer Engineering  and Birck Nanotechnology 
Center, \\ Purdue University, West Lafayette, IN 47907, USA
}%

\date{\today}

\begin{abstract}
We present the general analytical theory for Dyakonov surface waves at the interface of a 
biaxial anisotropic dielectric with an isotropic medium. We demonstrate that these surface waves can be divided into todo distinct classes, with qualitatively different spatial behavior. We obtain explicit expressions
for the Dyakonov waves dispersion and the parameter range for their  existence.
\end{abstract}

\maketitle


Electromagnetic surface waves, strongly localized near the interface of two different media, pay an important role in many areas of science and technology -- from optical microscopy \cite{SW_microscopy} and biosensing \cite{SW_biosensing} to nano-optical tweezing \cite{SW_tweezing} to photonic integrated circuits. \cite{SW_circuits} Electomagnetic surface waves are responsible for such phenomena as superlensing,\cite{Pendry2000,Haldane_arxiv} enhanced Raman scattering \cite{SW_Raman1,SW_Raman2} and extraordinary light transmission through subwavelength holes. \cite{Ebbesen} While there exists a number of different kinds of surface electormagnetic waves, such as e.g. surface plasmons at the interface of a metal and a dielectric, \cite{plasmonics} or  Tamm-Shockeley states \cite{Tamm,Shockley,Yeh1978} at  the boundary of a photonic crystal, \cite{Yablonovich,SJohn,Joannopoulos_book} a new class of surface electromagnetic modes has recently gained considerable attention. \cite{Marchevskii,Dyakonov1,Dyakonov2,Dyakonov3,SW_good_paper, SW_review,Dyakonov_experiment_PRL} These Dyakonov surface waves exist at the interface of an isotropic and anisotropic dielectric media. They can be supported by transparent optical materials, and thus do not suffer from the metallic absorption that plagues surface plasmons. \cite{ref:figure-of-merit} Compared to the Tamm-Shockley state, Dyakonov wave does not require any period patterning of the material forming the system, with the resulting light scattering due to the inevitable disorder as a result of an imperfect fabrication of such lattice.

The presence of Dyakonov waves at the isotopic-anisotropic interface has been firmly established in the experiment, \cite{Dyakonov_experiment_PRL} and a number of adequate theoretical methods exists for their quantitative description. \cite{Dyakonov1,Dyakonov2,SW_good_paper} However, due to the inevitable complexity of the boundary conditions at the interface of a fully-anisotropic dielectric the resulting theoretical description generally leads to a system of nonlinear equations that must be solved numerically. While this may be considered a straightforward task, Dyakonov waves
are usually extended over many wavelengths, \cite{SW_good_paper}  and are therefore close 
to the propagation wave threshold -- which makes the numerical solution more challenging. What is even more important, with the theoretical ``toolbox'' limited to numerical methods, the root-finding algorithm may even miss 
an entire class of possible solutions.

In this work, we present a complete analytical solution for the Dyakonov surface waves at the interface of an isotropic and a biaxial dielectric medium. We show that, depending on the magnitudes of the dielectric permittivity components in the system, the interface can simultaneously support two different classes of surface waves, with qualitatively different spatial behavior. 

\begin{figure}[hbt]
\includegraphics[width=3. in]{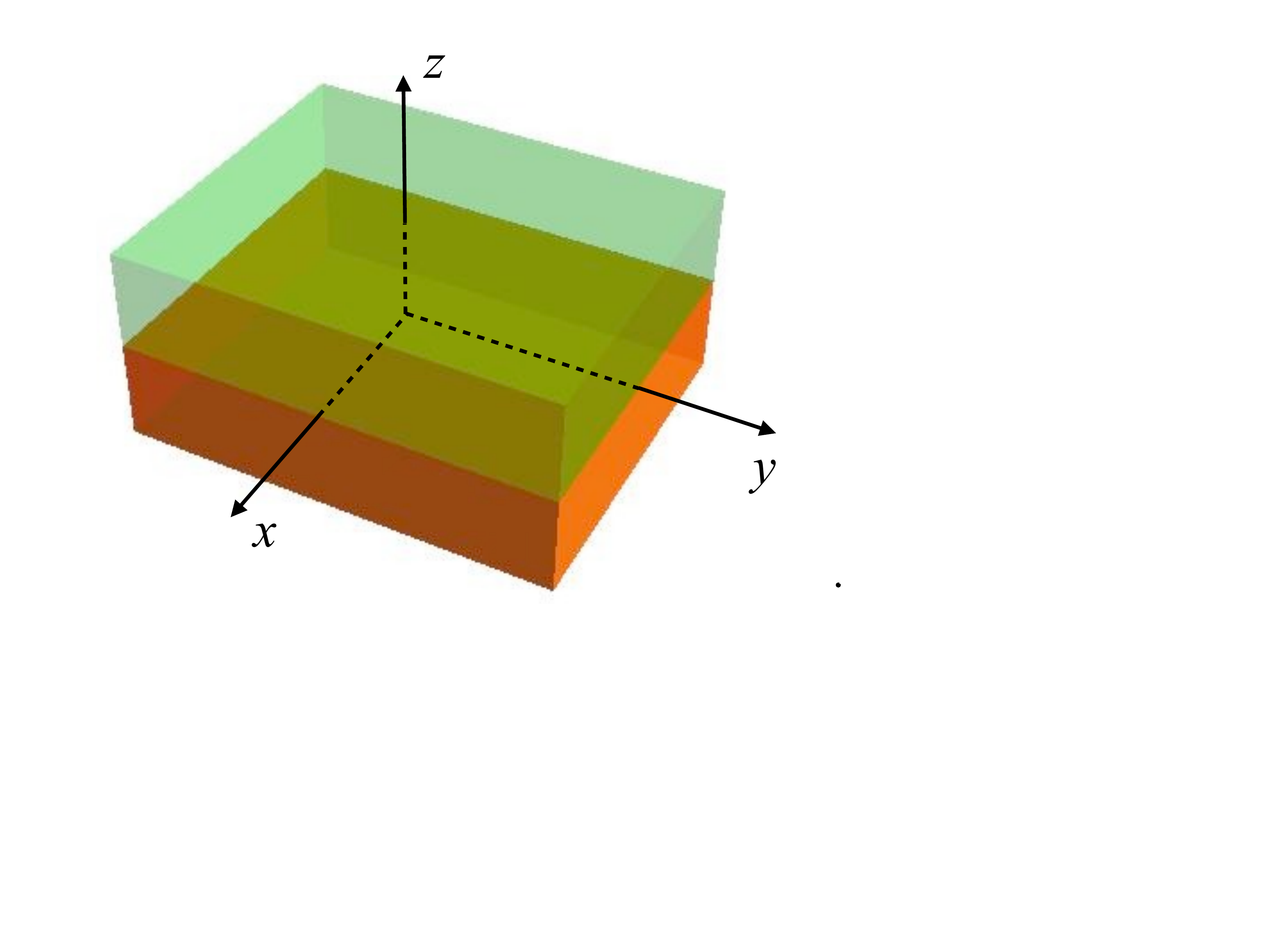}
\caption{\label{fig:schematics} 
The schematics of the coordinate system at the planar interface of a transparent  isotropic medium (orange) and biaxial anisotropic dielectric (green area).
}
\end{figure}

\section{The model}

We consider the interface of an isotopic dielectric with the permittivity $\epsilon_0$, with a biaxial anisotropic medium, with the permittivity tensor
\begin{eqnarray}
\epsilon & = & 
\left(
\begin{array}{ccc}
\epsilon_x & 0 & 0 \\
0 & \epsilon_y & 0 \\
0 & 0 & \epsilon_z 
\end{array}
\right).
\label{eq:eps_an}
\end{eqnarray}
We furthermore assume that one of the symmetry directions of the anisotropic crystal (which will be referred to as the axis $z$ in our coordinate system -- see Fig. \ref{fig:schematics}) is normal to the interface, as this is generally the case for a high-quality interface. While a non-orthogonal orientation of $\hat{\bf z}$ with respect to the plane of surface  is  possible, this would lead to a relatively high density of surface defects -- thus making the theory for surface waves at  a idea planar interface irrelevant for most practical application. For convenience, the coordinate system origin  $z = 0$ is chosen at the plane of the interface -- see Fig. \ref{fig:schematics}.

In this work, we focus on guided surface waves with the in-plane momenttum ${\bf q} \equiv \left( q_x, q_y\right)$,
\begin{eqnarray}
{\bf E}\left({\bf r}, t \right)& = & {\bf E}_{\bf q}\left(z\right)\cdot \exp\left(i q_x x + iq_y y - i \omega t\right), \label{eq:eq}\\
{\bf B}\left({\bf r}, t \right) & = & {\bf B}_{\bf q} \left(z\right)\cdot \exp\left(i q_x x + iq_y y - i \omega t\right), \label{eq:bq}
\end{eqnarray}
where
\begin{eqnarray}
E_{\bf q} \left(\left| z \right| \to \infty\right)  \to 0, \ B_{\bf q}\left(\left| z \right| \to \infty \right) \to 0
\end{eqnarray}

\begin{figure*}[hbt]
\includegraphics[width=6.5 in]{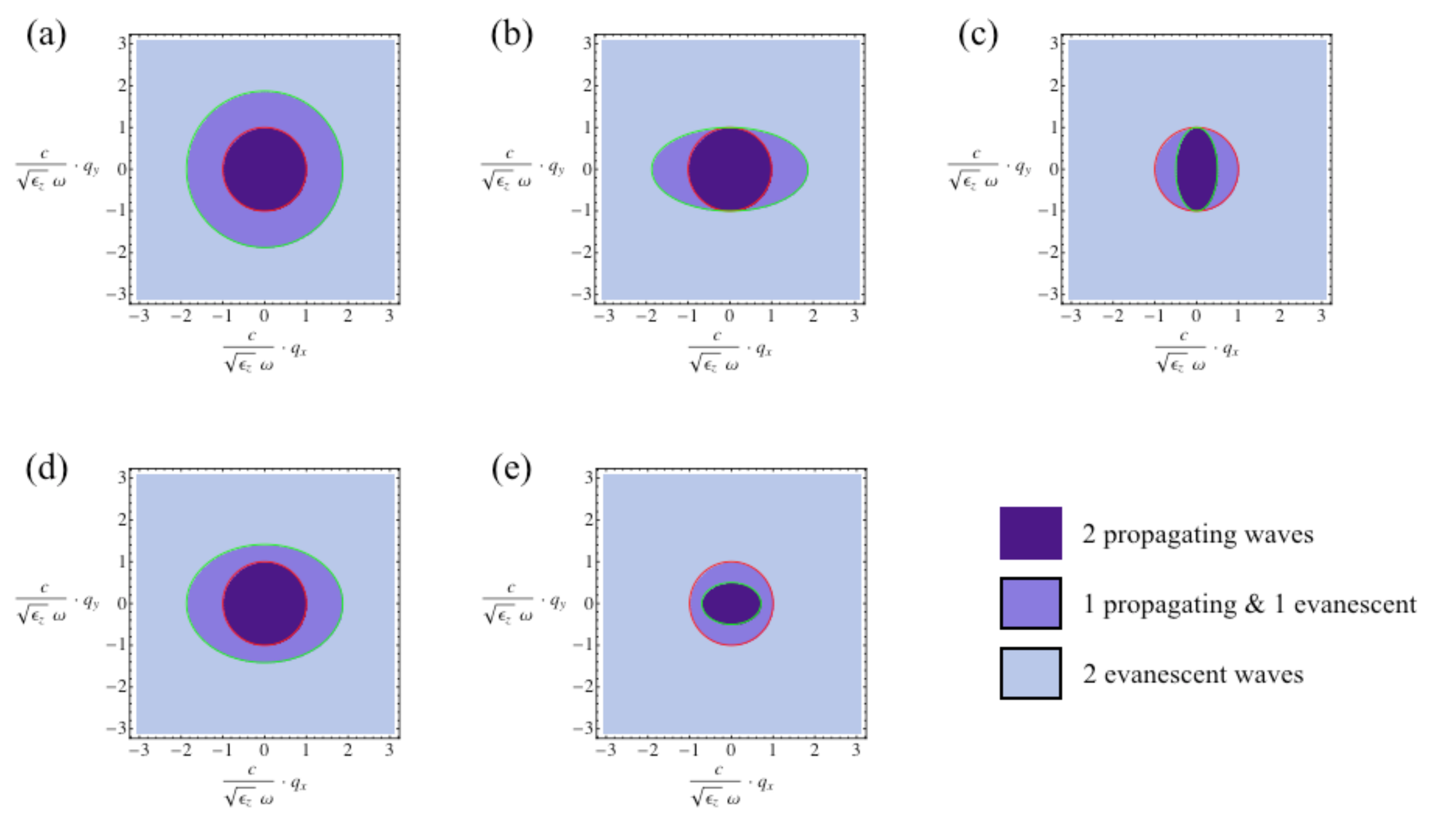}
\caption{\label{fig:no_ghosts} The ``phase space'' for the waves with ``in-plane'' momentum ${\bf q}$ (see Eqns. (\ref{eq:e_kp}), (\ref{eq:b_kp})) supported by an anisotropic dielectric. Panels (a), (b) and (c) correspond to a uniaxial dielectric, with $\epsilon_x = \epsilon_y > \epsilon_z$ (a), $\epsilon_x = \epsilon_z < \epsilon_y$ (b), and
$\epsilon_x = \epsilon_z > \epsilon_y$ (c) respectively. Panels (d) and (e) represent the case of a biaxial 
dielectric, correspondingly with $\epsilon_z < \epsilon_x < \epsilon_y$ (d), and $\epsilon_x < \epsilon_y < \epsilon_z$ (e). The  red  line corresponds to Eqn. (\ref{eq:boundary_z}), and the green line represents Eqn. (\ref{eq:boundary_xy}). The dielectric permittivity tensor components satisfy (a) $\epsilon_x = \epsilon_y = 3.5 \epsilon_z$, (b) $\epsilon_x = \epsilon_z = 3.5 \epsilon_y$, (c)  $\epsilon_x = \epsilon_z =   \epsilon_y / 4$, (d)  
$\epsilon_x = 2 \epsilon_z$ and $\epsilon_y = 3.5  \epsilon_z$, (e)  $\epsilon_x =  \epsilon_z / 4$ and $\epsilon_y =  \epsilon_z / 2$.
}
\end{figure*}

\section{Electromagnetic waves in a biaxial medium}

For an evanescent wave that decays away from the $z = 0$ interface, we have
\begin{eqnarray}
{\bf E}_{\bf q}\left( z\right) & = & {\bf e} \cdot  \exp\left( - \kappa z\right), \label{eq:e_kp} \\
{\bf B}_{\bf q}\left( z\right) & = & {\bf b}  \cdot \exp\left( - \kappa z\right), \label{eq:b_kp}
\end{eqnarray}
Note that for a complex $\kappa$, the expressions (\ref{eq:e_kp}), (\ref{eq:b_kp}) also describe the propagating waves in the medium.

Substituting (\ref{eq:eq}),(\ref{eq:bq}) with (\ref{eq:e_kp}), (\ref{eq:b_kp})   into Maxwell's equations, we obtain
\begin{eqnarray}
b_x & = & \frac{c}{\omega}\left( q_y e_z - i \kappa e_y\right), \label{eq:bx} \\
b_y & = & \frac{c}{\omega}\left( i \kappa e_x - q_x e_z\right), \label{eq:by} \\
b_z & = & \frac{c}{\omega}\left( q_x e_y -  q_y e_x\right), \label{eq:bz} 
\end{eqnarray}
and
\begin{eqnarray}
{\cal M}  
\left(
\begin{array}{c}
e_x \\
e_y \\
e_z
\end{array}
\right) & = &  0,
\label{eq:det_e}
\end{eqnarray} 
where
\begin{eqnarray}
{\cal M}&  \equiv & 
\left(
\begin{array}{ccc}
\Delta_x\left(\kappa\right)  & q_x q_y & i \kappa q_x \\
q_x q_y & \Delta_y\left(\kappa\right)  & i \kappa q_y \\
i \kappa q_x & i \kappa q_y & \Delta_z\left(\kappa\right) 
\end{array}
\right),
\end{eqnarray}
and
\begin{eqnarray}
\Delta_x\left(\kappa\right) & = & \epsilon_x \left(\frac{\omega}{c}\right)^2 - q_y^2 + \kappa^2, \\
\Delta_y\left(\kappa\right)  & = & \epsilon_y \left(\frac{\omega}{c}\right)^2 - q_x^2 + \kappa^2, \\
\Delta_z\left(\kappa\right)  & = & \epsilon_z \left(\frac{\omega}{c}\right)^2 - q_x^2 -q_y^2.
\end{eqnarray}
From  (\ref{eq:det_e}) we find the electrical field components in terms of the amplitude $a$
\begin{eqnarray}
e_x & =  &i \kappa q_x \left(q_y^2 - \Delta_y\left(\kappa\right) \right) \cdot a, \label{eq:ex} \\
e_y & = & i \kappa q_y \left(q_x^2 - \Delta_x\left(\kappa\right) \right) \cdot a, \label{eq:ey} \\
e_z & = & \left( \Delta_x\left(\kappa\right) \cdot  \Delta_y\left(\kappa\right)  - q_x^2 q_y^2 \right) \cdot a, 
\label{eq:ez}
\end{eqnarray}
which together with (\ref{eq:bx})-(\ref{eq:bz}) define the entire electromagnetic field ({\bf e}, {\bf b})  in (\ref{eq:e_kp}), (\ref{eq:b_kp}). 

Also, from Eqn. (\ref{eq:det_e}) we obtain
\begin{eqnarray}
{\rm det}\left[ {\cal M} \right] & = & 0,
\label{eq:det_M}
\end{eqnarray}
which yields
\begin{eqnarray}
  \epsilon_z \cdot \kappa^4  & + &  \bigg[  \left( \epsilon_x + \epsilon_y \right)  \cdot \left(\frac{\omega}{c}\right)^2  - \left(\epsilon_x + \epsilon_z\right)\cdot q_x^2  \nonumber \\
  & -& \left(\epsilon_y + \epsilon_z\right)\cdot q_y^2  \bigg]\cdot\kappa^2 
 +  \left(\epsilon_z \left(\frac{\omega}{c}\right)^2 - q_x^2 - q_y^2\right)  \nonumber \\
 & \times & \left(\epsilon_x \epsilon_y \left(\frac{\omega}{c}\right)^2  - \epsilon_x q_x^2 -  \epsilon_y q_y^2\right)  =  0. \label{eq:kappa_QE}
\end{eqnarray}
Eqn. (\ref{eq:kappa_QE}) is a quadratic equation for $\kappa^2$, with the straightforward solution
\begin{eqnarray}
\kappa_\pm^2 & = & \frac{1}{2} \left\{ \frac{\epsilon_x + \epsilon_z}{\epsilon_z} q_x^2 +   \frac{\epsilon_y+\epsilon_z }{\epsilon_z} q_y^2 -  \left( \epsilon_x + \epsilon_y \right)  \cdot \left(\frac{\omega}{c}\right)^2 \right. \nonumber \\
&  \pm & \left. \sqrt{D}\right\},
\label{eq:kappa_pm}
\end{eqnarray}
where the Discriminant
\begin{eqnarray}
D & = & \left[ \left(\epsilon_x - \epsilon_y\right) \left(\frac{\omega}{c}\right)^2 + \frac{\epsilon_z - \epsilon_x}{\epsilon_z} q_x^2 +   \frac{\epsilon_y - \epsilon_z }{\epsilon_z} q_y^2 \right]^2 \nonumber \\ & + &  4  \cdot \frac{\left(\epsilon_y - \epsilon_z \right)\cdot\left(\epsilon_y - \epsilon_z \right) }{\epsilon_z^2 } \ q_x^2 \ q_y^2.
\label{eq:discr}
\end{eqnarray}
When the Discriminant is positive, there are three distinct possibilities for the nature of the waves supported by the anisotropic dielectric. If the right-hand side of Eqn. (\ref{eq:kappa_pm}) is positive for  $\kappa_+^2$ and $\kappa_-^2$, both waves with the ``in-plane'' momentum ${\bf q} \equiv (q_x, q_y)$ are evanescent. In the opposite case, when the right-hand side of  Eqn. (\ref{eq:kappa_pm}) is negative in both cases, the corresponding two waves are propagating. Finally, when it's positive for one choice of the sign in  (\ref{eq:kappa_pm}) and negative for  the other, we find that for the given in-plane momentum ${\bf q}$  the dielectric interface supports one propagating and one evanescent wave.

As follows from Eqn. (\ref{eq:discr}), the Discriminant is  positive-definite (for any ${\bf q}$) in each of the following cases:
\begin{itemize}
\item{any uniaxial dielectric \\ ($\epsilon_x = \epsilon_y$ or $\epsilon_x = \epsilon_z$ or $\epsilon_y = \epsilon_z$),}
\item{$\epsilon_z < {\rm min}\left[\epsilon_x, \epsilon_y\right]$,}
\item{$\epsilon_z > {\rm max}\left[\epsilon_x, \epsilon_y\right]$.}
\end{itemize}
The boundaries that separate different portions of the $(q_x, q_y)$ phase space that respectively support only the propagating waves, or only the evanescent fields, or a mixture of evanescent and propagating waves, are given by
\begin{eqnarray}
q_x^2 + q_y^2 & = & \epsilon_z \left(\frac{\omega}{c}\right)^2,
\label{eq:boundary_z}
\end{eqnarray}
and
\begin{eqnarray}
\frac{q_x^2}{\epsilon_y}  + \frac{q_y^2}{\epsilon_x} & = & \left(\frac{\omega}{c}\right)^2,
\label{eq:boundary_xy}
\end{eqnarray}
This behavior is illustrated in Fig. \ref{fig:no_ghosts}.

However, if
\begin{eqnarray}
{\rm min}\left[\epsilon_x, \epsilon_y\right] < \epsilon_z < {\rm max}\left[\epsilon_x,\epsilon_y\right],
\label{eq:ghost_eps}
\end{eqnarray}
the Discriminant in Eqn. (\ref{eq:discr}) can, and does, for certain ranges of the values of $q_x$ and $q_y$, become negative. In this case, $\kappa_\pm$ is complex, with nonzero values for both its real and imaginary parts. These ``ghost waves'', recently described in Ref. \cite{EN_ghosts}, combine the oscillatory behavior of the propagating waves with the exponential decay characteristic of the evanescent fields, and represent the third class of the waves that can be supported by a transparent dielectric medium.

\begin{figure*}[hbt]
\includegraphics[width=6.5 in]{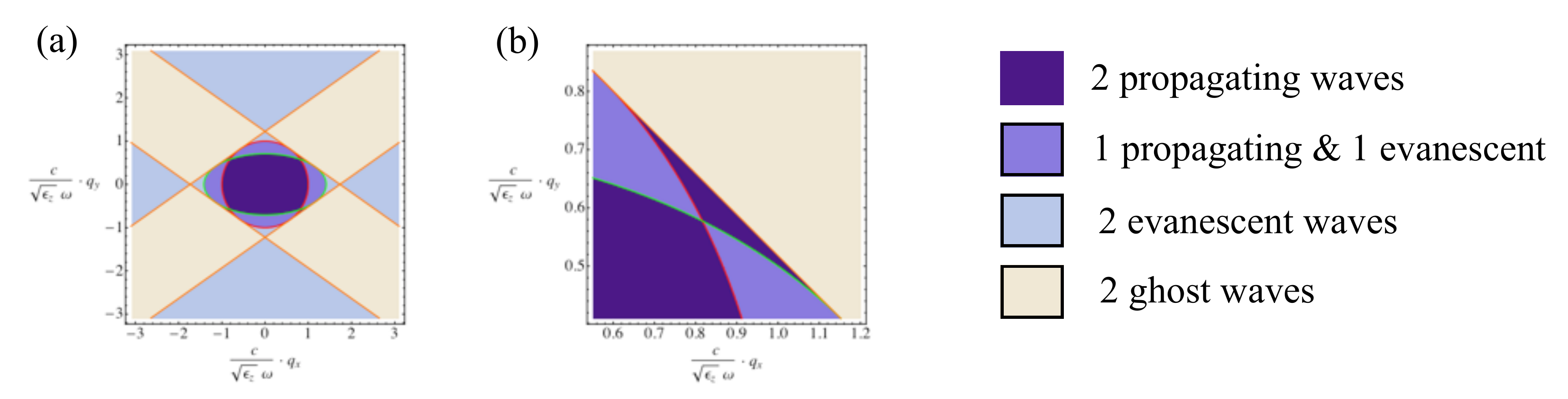}
\caption{\label{fig:yes_ghosts} The ``phase space'' for the waves with ``in-plane'' momentum ${\bf q}$ (see Eqns. (\ref{eq:e_kp}), (\ref{eq:b_kp})) supported by a biaxial anisotropic dielectric with $\epsilon_x < \epsilon_z < \epsilon_y$. Note the presence of the ghost waves in the regions bounded by four orange lines defined by 
Eqn. (\ref{eq:ghost_boundaries}). As in Fig. \ref{fig:no_ghosts}, the  red  line corresponds to Eqn. (\ref{eq:boundary_z}), and the green line represents Eqn. (\ref{eq:boundary_xy}). Panels (a) and (b) show the ``full" and the ``magnified" view of the phase space. Here $\epsilon_x /\epsilon_z = 0.5$ and $\epsilon_y/\epsilon_z = 2$.
}
\end{figure*}

When the inequality (\ref{eq:ghost_eps}) is satisfied, the boundaries of the portion of the $(q_x, q_y)$ phase space of the ghost modes are defined by the four equations
\begin{eqnarray}
\sqrt{\frac{\left| \epsilon_x - \epsilon_z\right|}{\epsilon_z}} q_x \pm \sqrt{\frac{\left| \epsilon_y - \epsilon_z\right|}{\epsilon_z}} q_y \pm \sqrt{\left| \epsilon_y - \epsilon_x\right|} \frac{\omega}{c} = 0, \ \ \ 
\label{eq:ghost_boundaries}
\end{eqnarray}
Fig. \ref{fig:yes_ghosts} shows the phase space of a biaxial anisotropic dielectric that supports ghosts waves. Note its nontrivial structure near the point corresponding to the intersection of the boundaries described by Eqns. (\ref{eq:boundary_z}) and (\ref{eq:boundary_xy}) in the magnified view of its panel (b).

\begin{figure}[hbt]
\includegraphics[width=3.5 in]{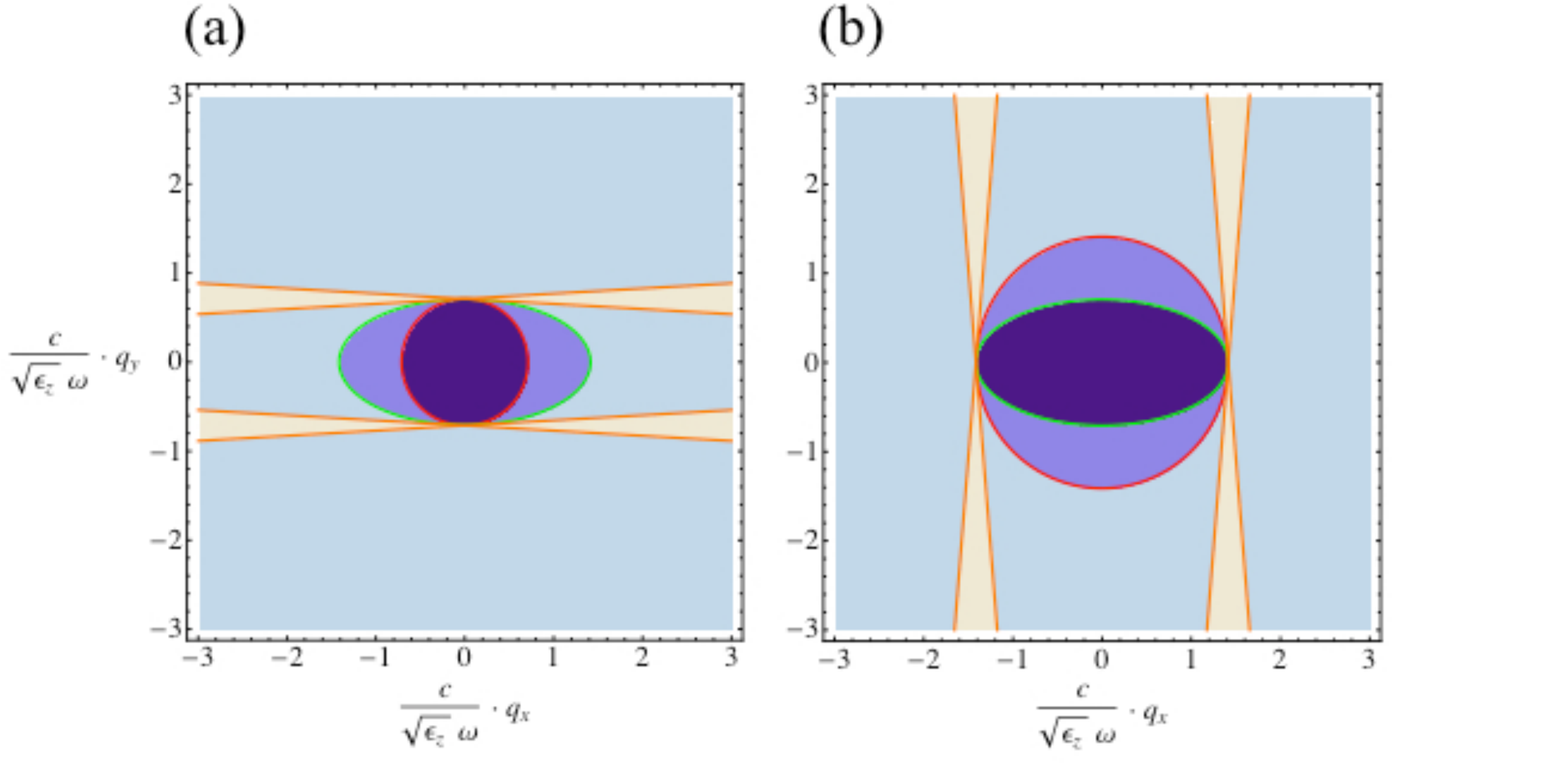}
\caption{\label{fig:collapse} The phase space for a biaxial anisotropic material for $\epsilon_z \to \epsilon_x$ (panel (a)) and $\epsilon_z \to \epsilon_y$ (panel (b)). In both cases $\epsilon_y / \epsilon_x  = 4$, while the $z$-component of the permittivity $\epsilon_z$ is such that $\epsilon_z/\epsilon_x = 1.01$ (a) and  $\epsilon_z/\epsilon_y = 0.99525$ (b). The phase space color code is that same as in  Figs. \ref{fig:no_ghosts} and \ref{fig:yes_ghosts}. Note the presence of relatively large ghost regions even though $\epsilon_z$ in both cases is within $1\%$ from its limiting values.
}
\end{figure}

When the permittivity $\epsilon_z$  in the normal-to-the-interface direction approaches the value of one of the  in-plane permittivities $\epsilon_x$ or $\epsilon_y$, the ghost regions in the phase space collapse to increasingly narrow strips parallel to either  the  $q_x$ (when $\epsilon_z \to \epsilon_x$) or $q_y$ (for  $\epsilon_z \to \epsilon_y$) axis. This ``collapse'' is however relatively slow, and substantial ghost regions are still present even when the permittivity is within 1\% of the critical value,  as seen in Fig. \ref{fig:collapse}.

\begin{figure}[hbt]
\includegraphics[width=3.5 in]{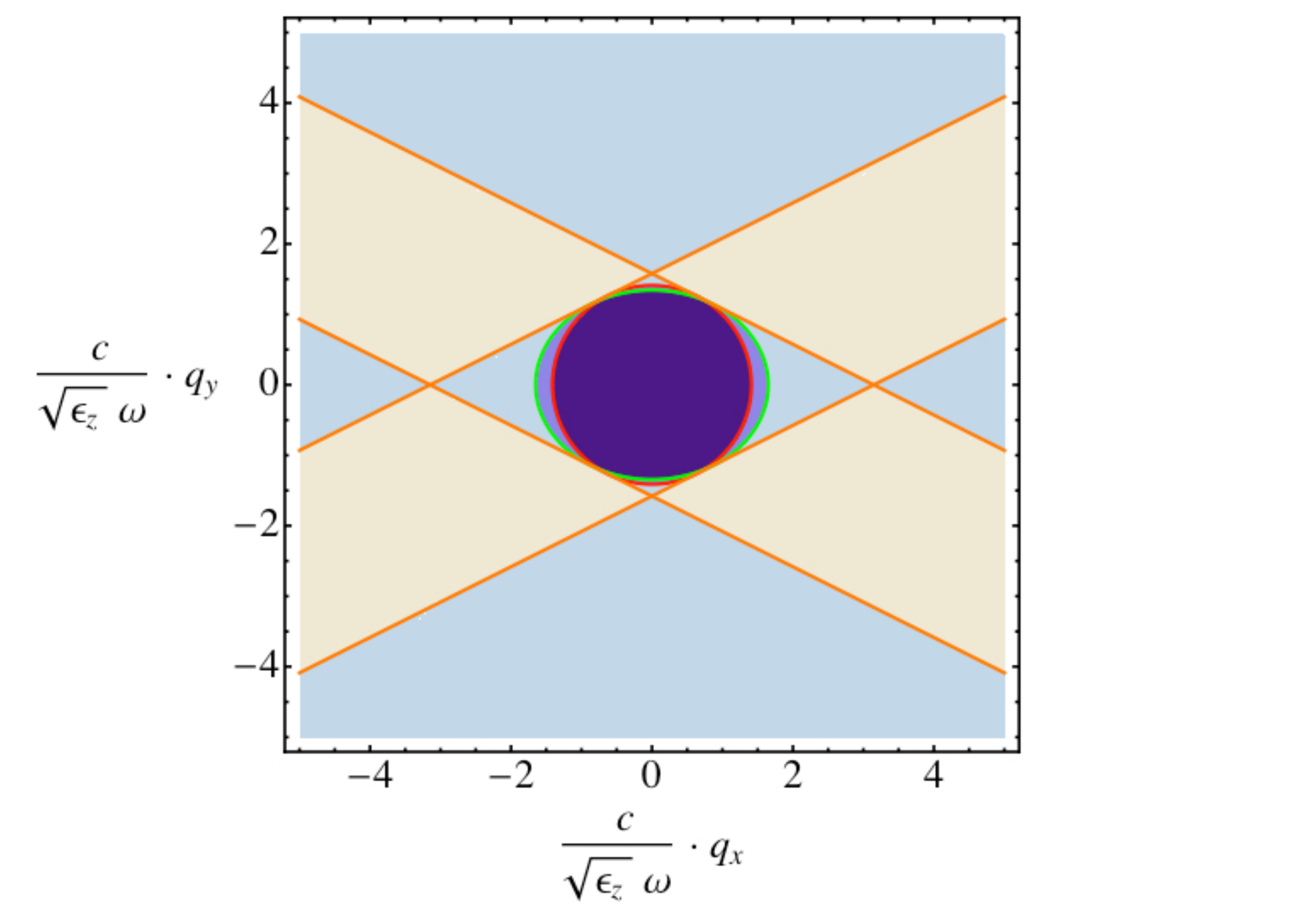}
\caption{\label{fig:sodium_nitrite} The phase space for sodium nitrite, with  $\epsilon_x = 1.806$, $\epsilon_y = 2.726$ and $\epsilon_z  = 1.991$.
}
\end{figure}

Most importantly, ghost regions show substantial presence in actual biaxial anisotropic crystals. This is illustrated in  Fig. \ref{fig:sodium_nitrite}, where we show the phase space for the  sodium nitrite $\rm NaNO_2$, with the dielectric permittivity tensor components \cite{moti} $\epsilon_x = 1.806$, $\epsilon_y = 2.726$ and $\epsilon_z  = 1.991$.    

While Eqns. (\ref{eq:bx}) - (\ref{eq:ez}) adequately describe the general  case of a dielectric crystal with arbitrary degree of anisotropy, the isotropic limit $\epsilon_x \to \epsilon_y  \to \epsilon_z \to \epsilon_0$ is singular, as here both $\kappa_+$ and $\kappa_-$ are identical,
\begin{eqnarray}
\kappa_+\left(\epsilon_x, \epsilon_y, \epsilon_z \to \epsilon_0\right) =  \kappa_-\left(\epsilon_x, \epsilon_y, \epsilon_z \to \epsilon_0\right) = \kappa_0,
\label{eq:kappa_limit}
\end{eqnarray} 
with
\begin{eqnarray}
\kappa_0 & = & q_x^2 + q_y^2 - \epsilon_0 \left(\frac{\omega}{c}\right)^2,
\label{eq:kappa0}
\end{eqnarray}
and direct substitution of (\ref{eq:kappa_limit}),(\ref{eq:kappa0}) into (\ref{eq:ex}), (\ref{eq:ey}), (\ref{eq:ez}) and  (\ref{eq:bx}), (\ref{eq:by}), (\ref{eq:bz}) yields
\begin{eqnarray}
e_x, e_y, e_z, b_x, b_y, b_z  & \to & 0\cdot a_0, 
\end{eqnarray}
with $a_0 \to \infty$. This uncertainty can be removed if we explicitly introduce $s-$ and $p-$ polarizations, correspondingly with $e_z^{(s)} = 0$ and $b_z^{(p)} = 0$:
\begin{eqnarray}
e_x^{(s)} & = & q_y \cdot a_s, \label{eq:e_s_x} \\
e_y^{(s)} & = & - q_x \cdot a_s, \\
e_z^{(s)} & = & 0, \\
b_x^{(s)} & = & \frac{i c \kappa_0}{\omega } \  q_x \cdot a_s, \\
b_y^{(s)} & = & \frac{i c \kappa_0}{\omega } \  q_y \cdot a_s, \\
b_z^{(s)} & = & - \frac{ c q^2 }{\omega }  \cdot a_s,
\end{eqnarray}  
and
\begin{eqnarray}
e_x^{(p)} & = & q_x \cdot a_p, \\
e_y^{(p)} & = & q_y \cdot a_p, \\
e_z^{(p)} & = & \frac{i q^2}{\kappa_0} \cdot a_p, \\
b_x^{(p)} & = & \frac{i \omega \epsilon_0}{c \kappa_0} \  q_y \cdot a_p, \\
b_y^{(p)} & = & - \frac{i \omega  \epsilon_0}{c \kappa_0} \  q_x \cdot a_p, \\
b_z^{(p)} & = & 0. \label{eq:b_p_z}
\end{eqnarray}  
Here \begin{eqnarray}
q \equiv \sqrt{q_x^2 + q_y^2},
\end{eqnarray}
while $a_s$ and $a_p$ are the scaled amplitudes of the $s-$ and $p-$polarized waves respectively.

\section{Dyakonov Wave}

Assuming that the interface at $z = 0$ separates transparent isotropic medium with the permittivity $\epsilon_0$ at $z < 0$ from biaxial anisotropic dielectric with the permittivity tensor (\ref{eq:eps_an}), for the guided surface wave with the in-plane momentum ${\bf q} = (q_x, q_y)$ we obtain
\begin{eqnarray}
{\bf E_q}\left(z\right) & = & \left\{
\begin{array}{cc}
\left( a_s {\bf e}_s + a_p {\bf e}_s \right) e^{  \kappa_0 z},  & z < 0 \\
a_+ {\bf e}_+ e^{ - \kappa_- z} +a_- {\bf e}_+ e^{ - \kappa_- z}, & z > 0
\end{array}
\right.
\label{eq:SW}
\end{eqnarray}
and
\begin{eqnarray}
{\bf B_q}\left(z\right) & = & \left\{
\begin{array}{cc}
\left( a_s {\bf b}_s + a_p {\bf b}_s \right) e^{  \kappa_0 z},  & z < 0 \\
b_+ {\bf e}_+ e^{ - \kappa_- z} +b_- {\bf e}_+ e^{ - \kappa_- z}, & z > 0
\end{array}
\right.
\end{eqnarray}
where (note the sign change $\kappa_0 \to -\kappa_0$ from (\ref{eq:e_s_x}) - (\ref{eq:b_p_z})  to (\ref{eq:e_s}) - (\ref{eq:b_p})  as the evanescent field for $z<0$ behaves as $\exp\left(+\kappa_0 z\right)$) 
\begin{eqnarray}
{\bf e}_s & = &  \left(q_y, \ -q_x, \ 0\right), \label{eq:e_s} \\
{\bf b}_s & = & - \frac{c}{\omega} \left( i \kappa_0 q_x,   i \kappa_0 q_y, \ q_x^2 + q_y^2 \right), \\
{\bf e}_p & = & \left(q_x, \ q_y, - \frac{i}{\kappa_0} \left(q_x^2 + q_y^2 \right)  \right),  \\
{\bf b}_p & = &  \frac{i \omega \epsilon_0}{c \kappa_0}  \  \left( - q_y, \   q_x, \ 0 \right), \label{eq:b_p}
\end{eqnarray}
and
\begin{eqnarray}
{\bf e}_\pm & = & \left( \  i \kappa_\pm \ q_x \left(q_y^2 - \Delta_y\left(\kappa_\pm \right) \right),  \ i \kappa_\pm  \ q_y \left(q_x^2 - \Delta_x\left(\kappa_\pm \right) \right), \right. 
\nonumber \\ 
& { }  & \left.  \ \    \Delta_x\left(\kappa_\pm \right) \cdot  \Delta_y\left(\kappa_\pm \right)  - q_x^2 q_y^2 \  \right), 
\\
{\bf b}_\pm & = &
\frac{\omega}{c}
 \left( \  q_y \left(\epsilon_y \Delta_x\left(\kappa_\pm\right) - \epsilon_x q_x^2\right), \right. \nonumber \\
 &  {-  } & \left. 
  q_x \left(\epsilon_x \Delta_y\left(\kappa_\pm\right) - \epsilon_y q_y^2\right), 
 i q_x q_y\kappa_\pm \   \left(\epsilon_y - \epsilon_x \right) \  \right). 
\end{eqnarray}
With non-magnetic ($\mu = 1$) dielectric materials at both sized of the interface, at $z = 0$ we have the continuity of all three components of the magnetic field ${\bf B_q}$, and the continuity of  $E_x$, $E_y$ and $D_z \equiv \epsilon_z E_z$. However, as follows from (\ref{eq:bz}), the continuity of both tangential components of the electric field immediately implies the continuity of $B_z$ Furthermore, since
\begin{eqnarray}
\epsilon_z E_z \propto \left[ {\rm curl} B \right]_z \propto q_x B_y - q_y B_x,
\end{eqnarray}
the continuity of $D_z = \epsilon_z E_z$ is a direct consequence of the continuity of the tangential magnetic field. Therefore, out of six boundary conditions here only four are actually independent, consistent with the four independent amplitudes $a_s$m $a_p$, $a_+$ and $a_-$.

Imposing the continuity of $E_x$, $E_y$, $\epsilon_z E_z$ and $\partial_z B_z \propto \left( q_x B_y + q_y B_x \right)$, we obtain:
\begin{eqnarray}
{\cal N} \left(
\begin{array}{c}
a_s \\
a_p \\
a_+ \\
a_-
\end{array}
\right) & = & 0,
\label{eq:N_a}
\end{eqnarray}
where the matrix ${\cal N}$ is defined as
\begin{eqnarray}
{\cal N} & = &
\left( 
\begin{array}{cccc}
\frac{ i q_y}{q_x}  & i &\kappa_+  \left(q_y^2 - \Delta_y^+\right) &  \kappa_- \left(q_y^2 - \Delta_y^-\right) \\
- \frac{i q_x}{q_y} & i &   \kappa_+ \left(q_x^2 - \Delta_x^+\right) &  \kappa_- \left(q_x^2 - \Delta_x^-\right) \\
0 & \frac{i q^2 \epsilon_0}{\kappa_0 \epsilon_z} & \Delta_x^+ \Delta_y^+ - q_x^2 q_y^2 & \Delta_x^- \Delta_y^- - q_x^2 q_y^2 \\
\frac{i q^2 \kappa_0}{q_x q_y} & 0 &  \left(\epsilon_y - \epsilon_x\right) \frac{\omega^2 \kappa_+^2}{c^2} & 
  \left(\epsilon_y - \epsilon_x\right) \frac{\omega^2\kappa_-^2}{c^2}
\end{array}
\right),
\nonumber \\ 
& { } & \ \ \
\label{eq:N}
\end{eqnarray}
with
\begin{eqnarray}
 \Delta_{x,y}^\pm  \equiv  \Delta_{x,y}\left(\kappa_\pm\right).
\end{eqnarray}
Introducing the new variable $\zeta_\pm$ coresponding to the $z$-components of the amplitudes of the electric field in the anisotropic material $\left({\bf e}_+\right)_z$ and $\left({\bf e}_-\right)_z$,
\begin{eqnarray}
\zeta_\pm & = & \left( \Delta_x^\pm  \Delta_y^\pm - q_x^2 q_y^2 \right) \cdot a_\pm,
\end{eqnarray}
from (\ref{eq:N_a}) and (\ref{eq:N}) we obtain
\begin{eqnarray}
{\cal P}\left(\omega; {\bf q} \right)  \ 
\left(
\begin{array}{c}
\zeta_+  \\ \zeta_-
\end{array}
\right)
& = & 0,
\end{eqnarray}
where the matrix ${\cal P}$ is defined by
\begin{eqnarray}
{\cal P}\left(\omega; {\bf q} \right)  & = & 
\left(
\begin{array}{cc}
\alpha_+ & \alpha_- \\
\beta_+ &  \beta_-
\end{array}
\right),
\end{eqnarray}
and
\begin{eqnarray}
\alpha_\pm & = & \frac{\epsilon_z}{\epsilon_0} + \frac{\kappa_\pm}{\kappa_0} \nonumber \\
& \times & 
\frac{\left(\frac{\omega}{c}\right)^2 \left(\epsilon_x q_y^2 + \epsilon_y q_x^2 \right) - q^2\left(q^2 - \kappa_\pm^2\right)}{ \Delta_x^\pm \Delta_y^\pm - q_x^2 q_y^2}, \\
\beta_\pm & = & \kappa_\pm \cdot 
\frac{\kappa_0 + \kappa_\pm}{ \Delta_x^\pm \Delta_y^\pm - q_x^2 q_y^2}.
\end{eqnarray}
The dispersion of the surface wave is then given by
\begin{eqnarray}
{\rm det} \left[ {\cal P}\left(\omega; {\bf q} \right) \right] & = & 0,
\end{eqnarray}
which yields
\begin{eqnarray}
& & \kappa_0 \left(\kappa_+ + \kappa_-\right) \cdot \left\{ \frac{\epsilon_x \epsilon_y}{\epsilon_0} \left( \left(\frac{\omega}{c}\right)^2 - \frac{q_x^2}{\epsilon_y} - \frac{q_y^2}{\epsilon_x}\right) - \kappa_+ \kappa_- \right\} \nonumber \\
& + & \kappa_+\kappa_- \left\{ 
\left(\epsilon_x + \epsilon_y \right) \left(\frac{\omega}{c}\right)^2 -  \frac{\epsilon_0 + \epsilon_x}{\epsilon_0} q_x^2 -
 \frac{\epsilon_0 + \epsilon_y}{\epsilon_0} q_y^2
\right\}  \nonumber \\
& + & \left\{
\frac{\epsilon_x \epsilon_y}{\epsilon_0} \kappa_0^2 \left(
\left(\frac{\omega}{c}\right)^2 - \frac{q_x^2}{\epsilon_y} - \frac{q_y^2}{\epsilon_x}
 \right) - \kappa_+^2 \kappa_-^2
\right\} = 0 .
\label{eq:SW1}
\end{eqnarray}
Eqn. (\ref{eq:SW}) uniquely defines the dispersion relation of the Dyakonov surface wave $\omega\left({\bf q}\right)$, and is the primary result of this section.

For a guided surface wave, all its components, in both the isotropic and anisotropic sides of the interface, must decay away from the boundary. For $z < 0$, this implies that
\begin{eqnarray}
q  > \sqrt{\epsilon_0} \  \frac{\omega}{c}.
\label{eq:limit0}
\end{eqnarray}
At the same time, in the anisotopic medium the waves with the in-plane momentum, ${\bf q}$ can belong to either the evanescent or ghost sub-classes -- see Section II. From Eqns. (\ref{eq:boundary_z}) and (\ref{eq:boundary_xy}) we therefore obtain 
\begin{eqnarray}
q  > \sqrt{\epsilon_z} \ \frac{\omega}{c},
\label{eq:limit1}
\end{eqnarray}
and
\begin{eqnarray}
\frac{q_x^2}{\epsilon_y} + \frac{q_y^2}{\epsilon_x} > \left(\frac{\omega}{c}\right)^2.
\label{eq:limit2}
\end{eqnarray}
Eqns. (\ref{eq:limit0}), (\ref{eq:limit1}) and (\ref{eq:limit2}) substantially reduce the range of the momentum and frequency that needs to be explored in the numerical solution of Eqn. (\ref{eq:SW1}). Furthermore, as shown in Ref. \cite{SW_good_paper} (see also Appendix A), the Dyakonov surface wave only exists when
\begin{eqnarray}
{\rm min}\left(\epsilon_x, \epsilon_y\right) \leq \epsilon_z < \epsilon_0 < {\rm max}\left(\epsilon_x, \epsilon_y\right).
\label{eq:eps_bounds}
\end{eqnarray}

While the numerical solution of Eqn. (\ref{eq:SW}) is generally straightforward, for small - to - moderate anisotropy, the surface waves are known \cite{SW_review,SW_good_paper}  to be relatively weakly guided,
\begin{eqnarray}
\kappa_0  \ll \omega/c,
\end{eqnarray} 
which turns numerical root-finding into a challenging numerical problem \cite{SW_good_paper}. In the next section we will therefore develop the method for the analytical solution of Eqn. (\ref{eq:SW1}).

\section{Analytical Solution for the surface wave dispersion}

Despite its relative complexity,  Eqn. (\ref{eq:SW1}) is not transcendental, but only contains algebraic functions. As a result, it can be reduced to a polynomial equation. Furthermore, as we show in the present section, the resulting polynomial equation is of the 4th order, and therefore allows a complete analytical solution.

Choosing the $y$-direction at the one corresponding to the largest permittivity in the plane of the interface, 
\begin{eqnarray}
\epsilon_y > \epsilon_x,
\label{eq:constraintsXY}
\end{eqnarray}
 we introduce the new variable 
\begin{eqnarray}
u & = & \frac{\epsilon_x \epsilon_y}{\epsilon_0} \left(\frac{q_x^2}{\epsilon_y} + \frac{q_y^2}{\epsilon_x} - \left( \frac{\omega}{c} \right)^2 \right).
\label{eq:u}
\end{eqnarray}
Note that, as follows from (\ref{eq:limit2}), $u> 0$. Then
\begin{eqnarray}
\kappa_+^2 \kappa_-^2 & = & \frac{\epsilon_0}{\epsilon_z}  \left( q^2 - \epsilon_z \left(\frac{\omega}{c}\right)^2  \right) \cdot u,
\label{eq:kpkp2}
\end{eqnarray}
and
\begin{eqnarray}
\kappa_+ + \kappa_- & = & \left[ 
q^2 +\frac{\epsilon_0}{\epsilon_z}u -  \left( \epsilon_x + \epsilon_y - \frac{\epsilon_x \epsilon_y}{\epsilon_z} \right) \left(\frac{\omega}{c}\right)^2\right. \nonumber \\
& + & \left. 2 \  \sqrt{\frac{\epsilon_0}{\epsilon_z} \left(\epsilon_z \left(\frac{\omega}{c}\right)^2 - q^2 \right) \cdot u} \ 
\right]^{1/2}
\end{eqnarray}
We can then express Eqn. (\ref{eq:SW1}) as 
\begin{eqnarray}
\kappa_0 \left(\kappa_+ + \kappa_-\right) \cdot \left( u + \kappa_+\kappa_-\right) & = & \hat{A} \ \kappa_+ \kappa_- + \hat{B},
\label{eq:ABrhs}
\end{eqnarray}
where
\begin{eqnarray}
\hat{A} & = &  \left( \epsilon_x + \epsilon_y - \frac{\epsilon_x \epsilon_y}{\epsilon_z} \right) \left(\frac{\omega}{c}\right)^2 - q^2 - \frac{\epsilon_0}{\epsilon_z} \cdot u,
\end{eqnarray}
and
\begin{eqnarray}
\hat{B} & = & - \kappa_0^2 u - \kappa_+^2 \kappa_-^2.
\end{eqnarray}
We then square both sides of Eqn. (\ref{eq:AB}), which yields
\begin{eqnarray}
\hat{G} \ \kappa_+ \kappa_- & = &  \hat{F} ,
\label{eq:GF} 
\end{eqnarray}
where
\begin{eqnarray}
\hat{G} & = &  \epsilon_0  \left\{ \left(\frac{\epsilon_x\epsilon_y}{\epsilon_0} - \epsilon_x - \epsilon_y + \epsilon_0 \right) \left(\frac{\omega}{c}\right)^2 \right. \nonumber \\
& + &  \left[
\left(\frac{\epsilon_x}{\epsilon_0} + \frac{\epsilon_y - \epsilon_0}{\epsilon_z} -\frac{\epsilon_x \epsilon_y}{\epsilon_0^2} \right) q_x^2 \right.
\nonumber \\
&  + &
\left. \left. 
\left(\frac{\epsilon_y}{\epsilon_0} + \frac{\epsilon_x - \epsilon_0}{\epsilon_z} -\frac{\epsilon_x \epsilon_y}{\epsilon_0^2} \right) q_y^2 
\right]  
\right\}  \\
 & = & 
 2 \epsilon_0 \left\{ \left(1 - \frac{\epsilon_0}{\epsilon_z} \right) \cdot u + \kappa_0^2 \left( 
 \frac{\epsilon_x + \epsilon_y - \epsilon_0}{\epsilon_z} - \frac{\epsilon_x \epsilon_y}{\epsilon_0^2} \right) \right. 
 \nonumber \\
 &  + & \left.  
 \left( \frac{1}{\epsilon_0} - \frac{1}{\epsilon_z} \right)    
 \left( \epsilon_0^2 - \epsilon_0 \left(\epsilon_x + \epsilon_y \right) + \epsilon_x \epsilon_y \right)  \left(\frac{\omega}{c}\right)^2
 \right\}
\end{eqnarray}
and
\begin{eqnarray}
\hat{F} & = & - \epsilon_0 \left(1 - \frac{\epsilon_0}{\epsilon_z}\right) \left\{ \left(\epsilon_2 \left(\frac{\omega}{c}\right)^2 - u\right)^2 \right. \nonumber \\
&  & - \left. \epsilon_0 \left(1 - \frac{\epsilon_0}{\epsilon_z}\right) \left(\frac{\omega}{c}\right)^2  u \right\}
\nonumber \\
& + & \kappa_0^2 \left\{ \epsilon_0 \left( \frac{\epsilon_2^2}{\epsilon_z} + \left(1 - \frac{\epsilon_0}{\epsilon_z} \right) \left(\epsilon_1 + 2 \epsilon_2 \right)\right) \left(\frac{\omega}{c}\right)^2  \right . \nonumber \\
& & - \left.  \left(\epsilon_1 + \frac{\epsilon_0}{\epsilon_z} \left(\epsilon_0 + 2 \epsilon_2\right) - \frac{\epsilon_0^3}{\epsilon_z^2} \right) u \right\}
\nonumber \\
& - & \kappa_0^4 \cdot \left\{  \frac{\epsilon_0}{\epsilon_z} \left(\epsilon_1 + 2 \epsilon_2 \right) \right\},
\end{eqnarray}
with
\begin{eqnarray}
\epsilon_1 & = &
\epsilon_ 0 - \epsilon_x - \epsilon_y + \frac{\epsilon_x \epsilon_y}{\epsilon_z},
\label{eq:eps1} 
\end{eqnarray}
and
\begin{eqnarray}
\epsilon_2 & = & \epsilon_x + \epsilon_y - \epsilon_0 - \frac{\epsilon_x \epsilon_y}{\epsilon_0}
 = \frac{ \left(\epsilon_0 - \epsilon_x \right) \left(\epsilon_y - \epsilon_0 \right)}{\epsilon_0}
\label{eq:eps2}
\end{eqnarray}

\begin{figure}[hbt]
\includegraphics[width=3.5 in]{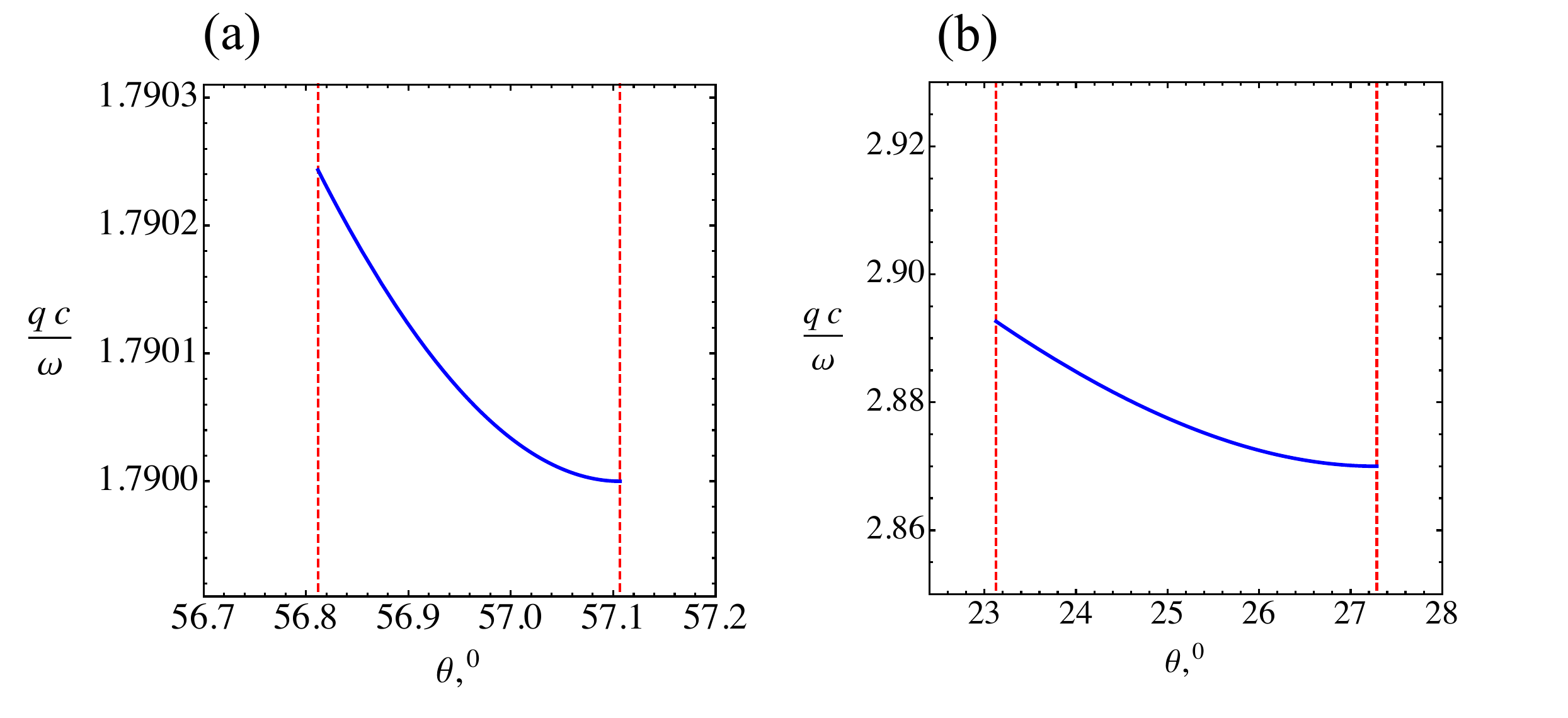}
\caption{\label{fig:SW1}The in-plane wavenumber of the Dyakonov surface wave (in units of $\omega/c$) vs. the propagation direction angle $\theta$ in biaxial materials, for the interfaces of (a) potassium titanyl phosphate (KTP) with  aluminium oxynitride (AlON), and (b) arsenic trisulfide with aluminum arsenide. The corresponding refractive indices are \cite{Weber} as follows. KTP : $n_x = 1.7614$,  $n_y  = 1.8636$, $n_z = 1.7704$;  ALON : $n_0 = 1.79$;  arsenic trisulfide : $n_x = 2.4$, $n_y = 3.02$, $n_z =   2.81$;  aluminum arsenide : $n_0 = 2.87$.
}
\end{figure}

Note that, in addition to the solutions of the original equation (\ref{eq:SW1}), the new Eqn. (\ref{eq:GF}) contains spurious roots corresponding to $\hat{A} \kappa_+ \kappa_- + \hat{B} < 0$. We therefore need to constrain the solutions of  (\ref{eq:GF}) with the inequality
\begin{eqnarray}
\hat{A} \  \kappa_+ \kappa_- + \hat{B} \ & > & 0.
\label{eq:AB}
\end{eqnarray}
Together, Eqns. (\ref{eq:GF}) and (\ref{eq:AB}) are equivalent to the original equation (\ref{eq:SW1}). 

Since $u > 0$ and $ q > \sqrt{\epsilon_z} \omega/c$ (see Eqns. (\ref{eq:boundary_z}),(\ref{eq:boundary_xy}) and (\ref{eq:u})), from Eqn. (\ref{eq:kpkp2}) we find
\begin{eqnarray}
\kappa_+  \kappa_- & = &\kappa_z \cdot  \sqrt{ \frac{\epsilon_0}{\epsilon_z} \cdot u} \  ,
\label{eq:kpkp}
\end{eqnarray}
where
\begin{eqnarray}
\kappa_z & \equiv & \sqrt{q^2 -   \epsilon_z }.
\end{eqnarray} 
Substituting (\ref{eq:kpkp}) into (\ref{eq:GF}), we obtain
\begin{eqnarray}
 \kappa_z  \cdot \hat{G} \cdot \sqrt{\frac{\epsilon_0}{\epsilon_z} \cdot u }& = & \hat{F}.
\label{eq:GF2}
\end{eqnarray}
Introducing the new dimensionless variable
\begin{eqnarray}
\chi & \equiv & \frac{c}{\omega} \kappa_z,
\end{eqnarray}
we can express Eqn. (\ref{eq:GF2}) in the form
\begin{eqnarray}
a_4\cdot\chi^4 +a_3\cdot\chi^3 +a_2\cdot\chi^2 +a_1\cdot\chi+a_0 & = & 0,
\label{eq:poly1}
\end{eqnarray}
where
\begin{eqnarray}
a_4 & = & \frac{\epsilon_0}{\epsilon_z} \left(\epsilon_1 + 2 \ \epsilon_2\right), \\
a_3 & = & \sqrt{\frac{\epsilon_x \epsilon_y}{\epsilon_0} \hat{u}} \cdot 2 \epsilon_0 \left(\frac{\epsilon_x + \epsilon_y - \epsilon_0}{\epsilon_z}   - \frac{\epsilon_x \epsilon_y}{\epsilon_0^2}\right), \\
a_2 & = & - \frac{\epsilon_0 \epsilon_2^2}{\epsilon_z} + \epsilon_0 \left(\epsilon_1 + 2 \epsilon_2\right) \cdot \left(1 - \frac{\epsilon_0}{\epsilon_z}\right) \nonumber \\
& + & \hat{u}\cdot \frac{\epsilon_x \epsilon_y}{\epsilon_0} \left( \epsilon_1 + 2 \cdot \frac{\epsilon_0 \epsilon_2}{\epsilon_z} + \frac{\epsilon_0^2}{\epsilon_z} \left(1 - \frac{\epsilon_0}{\epsilon_z}\right) \right), \\
a_1 & = & 2 \left(1 - \frac{\epsilon_0}{\epsilon_z}\right) \sqrt{\frac{\epsilon_x^3 \epsilon_y^3}{\epsilon_z}\hat{u} } \cdot \left(1 - \frac{\epsilon_z}{\epsilon_0} + \hat{u} \right), \\
a_0 & = & \left(1 - \frac{\epsilon_0}{\epsilon_z}\right) \cdot \frac{\epsilon_x \epsilon_y}{\epsilon_0}\cdot\hat{u}\cdot\left(\epsilon_1 \epsilon_z +  \epsilon_x \epsilon_y \hat{u} \right),
\end{eqnarray}
and
\begin{eqnarray}
\hat{u} & \equiv & \frac{\epsilon_0}{\epsilon_x \epsilon_y} \left(\frac{c}{\omega}\right)^2  \cdot u = 
 \left(\frac{c}{\omega}\right)^2\left( \frac{q_x^2}{\epsilon_y} + \frac{q_y^2}{\epsilon_x} \right) - 1.
 \label{eq:hatu}
\end{eqnarray}
The expression (\ref{eq:poly1}) is a quartic equation for $\chi$, and allows an immediate analytical solution via the Ferrari formula,~\cite{Ferrari} so that
\begin{eqnarray}
\chi & = & {\cal F}\left(\hat{u}; \epsilon_0, \epsilon_x, \epsilon_y, \epsilon_z\right).
\label{eq:chi_F}
\end{eqnarray}
Then, introducing the polar angle $\theta$ that defines the direction of the in-plane momentum ${\bf q}$,
\begin{eqnarray}
q_x & = q \cdot \cos\theta, \\
q_y & = q \cdot \sin\theta, 
\end{eqnarray}
from (\ref{eq:u}) and (\ref{eq:chi_F}) we obtain
\begin{eqnarray}
\frac{\omega}{c} & = & \frac{q}{\sqrt{\epsilon_z + {\cal F}^2\left(\hat{u}\right)}}, \label{eq:wq} \\
\sin\theta & = & \pm \sqrt{\frac{\epsilon_x \epsilon_y}{\left| \epsilon_y - \epsilon_x \right|} \cdot 
\left( \frac{\hat{u} + 1}{\epsilon_z + {\cal F}^2\left(\hat{u}\right)} - \frac{1}{\epsilon_y} \right)},
\label{eq:th}
\end{eqnarray}
which parametrically defines the function $\omega\left( q, \theta \right)$. 

\begin{figure}[hbt]
\includegraphics[width=3.5 in]{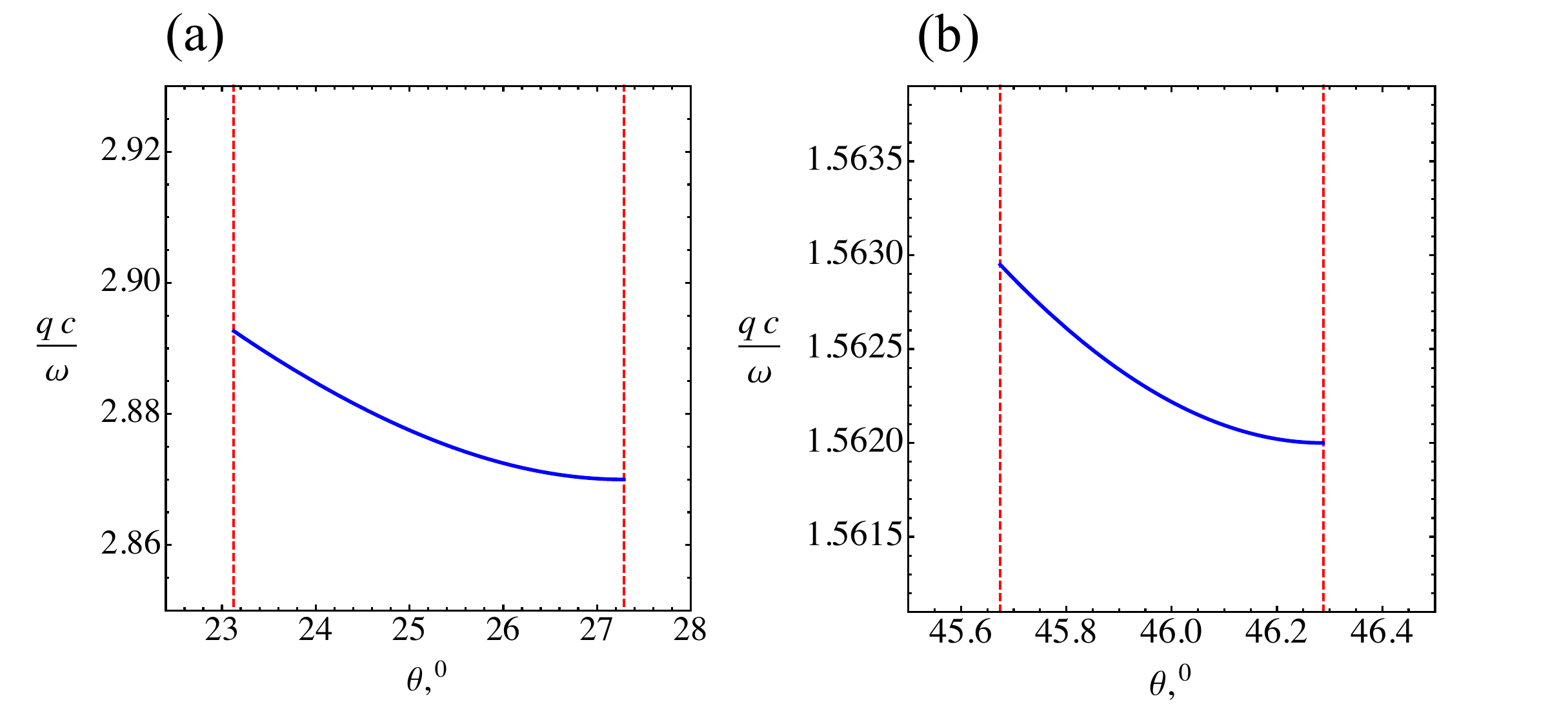}
\caption{\label{fig:SW2}
The in-plane wavenumber of the Dyakonov surface wave (in units of $\omega/c$) vs. the propagation direction angle $\theta$ in uniaxial materials, for the interfaces of (a) calcite with ${\rm CdF_2}$, and (b) lithium niobate  and ${\rm KTaO_3}$. The corresponding refractive indices are \cite{Weber} as follows. Calcite : $n_x = n_z  = 1.486$, $n_y = 1.658$; ${\rm CdF_2}$  : $n_0 = 1.562$;   lithium niobate  : $n_x =  n_z =  2.156$, $n_y =   2.232$;  ${\rm KTaO_3}$: $n_0 = 2.2$.
}
\end{figure}

In general, a quartic equation like (\ref{eq:poly1}) has four distinct roots. However, in our case $\chi$ should satisfy a number of additional constraints. Aside from being a positive real quantity, it must also exceed the value of $\sqrt{\epsilon_0 - \epsilon_z}$,
\begin{eqnarray}
\chi & > & \sqrt{\epsilon_0 - \epsilon_z},
\label{eq:chi_c}
\end{eqnarray}
since decay of the surface wave away from the interface implies
\begin{eqnarray}
\kappa_0 = \frac{\omega}{c} \sqrt{\chi^2 + \epsilon_z - \epsilon_0} > 0.
\end{eqnarray}

\begin{figure*}[hbt]
\includegraphics[width=6.5 in]{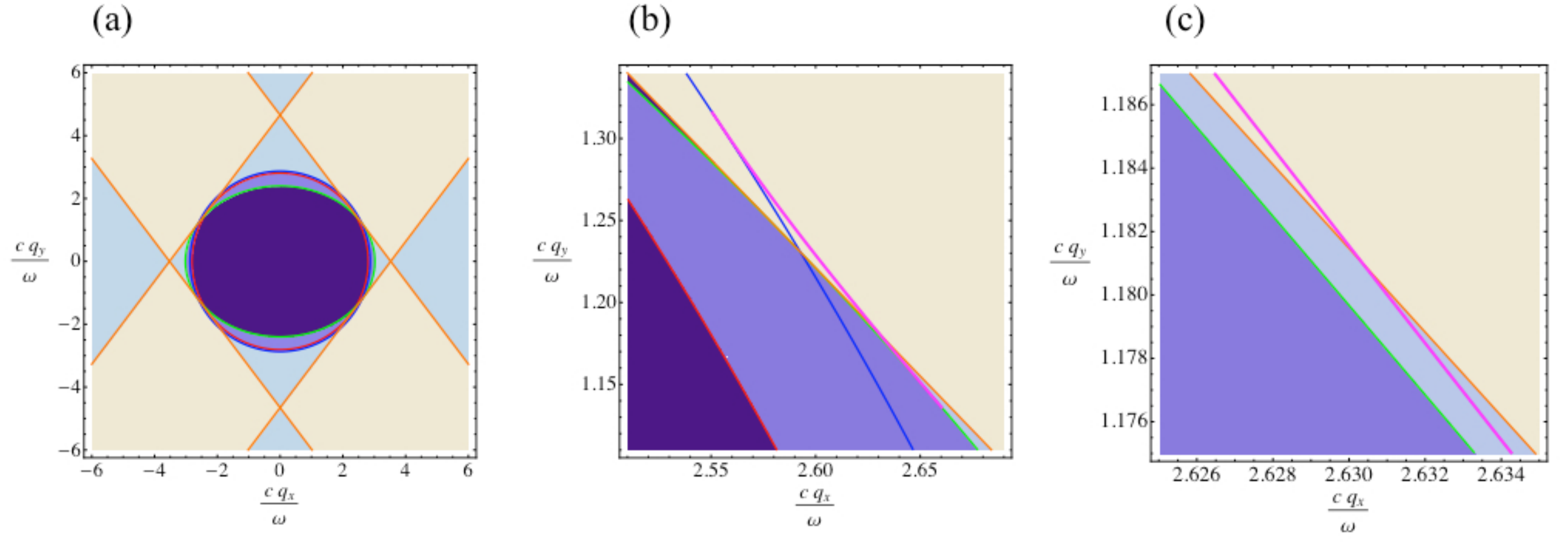}
\caption{\label{fig:projection} The Dyakonov surface wave in $q_x, q_y)$ coordinates for the aluminum arsenide - arsenic trisulfide interface. The surface wave is shown by the magenta line. The  red  line corresponds to Eqn. (\ref{eq:boundary_z}),  the green line represents Eqn. (\ref{eq:boundary_xy}),  the blue line corresponds to Eqn. (\ref{eq:limit0}), and the orange lines show the ghost region boundaries from Eqn. (\ref{eq:ghost_boundaries}). The phase space color code is the same as in Figs. 1 - 4. Panels (b) and (c) show the magnified portions of the phase space that supports the Dyakonov surface wave. Note that, as clearly seen in panel (c), the surface wave is supported by both the evanescent and the ghost regions of the phase space.
}
\end{figure*}

As we prove in Appendix B, Eqn. (\ref{eq:poly1}) only has no more than a single real positive solution that satisfies
(\ref{eq:chi_c}), so there is no ambiguity of choosing the correct root. We therefore obtain
\begin{eqnarray}
{\cal F}  & = & - \frac{a_3}{4 a_4} + s_1 S + \frac{s_2}{2} \sqrt{- 4\cdot S^2 - 2 \hat{p} -\frac{s_1\cdot \hat{q}}{S}}, 
\label{eq:F} 
\end{eqnarray}
where
\begin{eqnarray}
\hat{p} & = & \frac{a_2}{a_4} - \frac{3}{8}\cdot\frac{a_3^2}{a_4^2}, \\
\hat{q} & = & \frac{a_3^3 - 4 a_2 a_3 a_4 + 8 a_1 a_4^2  }{8 a_4^3}, \\
S & = & \frac{1}{2} \cdot \sqrt{- \frac{2}{3} \hat{p} + \frac{1}{3 a_4} \left( Q + \frac{\Delta_0}{Q}\right)}, \\
Q & = & \sqrt[3]{\frac{\Delta_1 + \sqrt{\Delta_1^2 - 4\cdot\Delta_0^3}}{2}}, \\
\Delta_0 & = & a_2^2 - 3 a_1 a_2 + 12 a_0 a_4, \\
\Delta_1 & = & 2 a_2^3 - 9 a_1 a_2 a_3 + 27 a_0 a_3^2 + 27 a_1^2 a_4 \nonumber \\
& - & 72 a_0 a_2 a_4, \\
s_{1,2} & = & \pm 1. \label{eq:s12}
\end{eqnarray}

While the choice of $s_1$ and $s_2$ in Eqn. (\ref{eq:s12}) that leads to a positive real root that satisfies Eqn. (\ref{eq:chi_c}), is unique, such a solution only exist in a limited range of angles $\theta$. Furthremore, the resulting solution must be tested against the inequality (\ref{eq:AB}) to remove the spurious roots. As a result, for the angular range of $\theta$ that supports the Dyakonov surface wave, we obtain (see Appendix C)  :
\begin{eqnarray}
\theta_1 < \left|  \theta \right|  < \theta_2, \ \ {\rm or} \ \ \pi - \theta_2 < \left|  \theta \right|  < \pi - \theta_1, 
\label{eq:range}
\end{eqnarray}
where, assuming $\epsilon_y > \epsilon_x$, 
\begin{eqnarray}
\theta_1 & = & {\rm arcsin}\left[ \left(\frac{\epsilon_x \epsilon_y}{\epsilon_y - \epsilon_x} \right. \right. \nonumber \\
& \times & \left. \left. 
\left( \frac{\epsilon_1 + 2 \epsilon_2}{\left(\epsilon_1 + 2 \epsilon_2\right) + \epsilon_2^2} - \frac{1}{\epsilon_y}\right) \right)^{1/2}  \right]
\label{eq:theta1}
 \end{eqnarray}
 and
 \begin{eqnarray}
 \theta_2 & = & {\rm arcsin}\left[\left( \left( 1 - 
 \sqrt{1 +  \frac{4 \epsilon_z \left( \epsilon_0 - \epsilon_x \right)\left( \epsilon_y - \epsilon_0 \right)}{\epsilon_0^2 \left(\epsilon_0 - \epsilon_z\right)} }\ \right) \right. \right. \nonumber \\
  & \times & \left. \left. 
   \frac{\epsilon_0}{2 \epsilon_z} \frac{\epsilon_0 - \epsilon_z}{\epsilon_y - \epsilon_x}  + 
  \frac{\epsilon_y - \epsilon_0}{\epsilon_y - \epsilon_x}  
 \right)^{1/2} \ \right]. 
 \label{eq:theta2}
\end{eqnarray}
Here, $\theta_1$ and $\theta_2$ correspond to $\kappa_- = 0$ and $\kappa_0 = 0$ respectively. At the same time, $\theta_1$ corresponds to the boundary of the inequality (\ref{eq:limit2}), while $\theta_2$ represents the ``edge'' of the inequality (\ref{eq:AB}) -- see Appendix C. Within the angle range (\ref{eq:range}) for any direction $\theta$ and the frequency $\omega$, there is one and only one surface wave, described by the parametric equations (\ref{eq:wq}), (\ref{eq:th})  with the function  ${\cal F}\left(\hat{u}, \epsilon_0, \epsilon_x, \epsilon_y, \epsilon_z\right)$ from  Eqn. (\ref{eq:F}), while for any angle outside this range, there is no surface wave.

In Fig. \ref{fig:SW1} we plot the surface wave dispersion for the interface of potassium titanyl phosphate (KTP) and aluminium oxynitride (AlON) (panel (a)), and arsenic trisulfide with aluminum arsenide (panel (b)). The results of the present work can also be applied to uniaxial materials, as illustrated in Fig. \ref{fig:SW2} for calcite and ${\rm CdF_2}$ (panel (a)), and lithium niobate (${\rm LiNbO_3}$) and ${\rm KTaO_3}$ (panel (b)).

Following Ref. \cite{SW_good_paper}, it is also instructive to project the surface wave dispersion onto the wavevector space $(q_x, q_y)$ that we studied in Section II. In Fig. \ref{fig:projection} we show this projection for the surface wave at the interface of isotropic aluminum arsenide and biaxial arsenic trisulfide. As expected, the magenta curve that represents the Dyakonov surface wave, terminates at the boundaries corresponding to 
$\kappa_0 = 0$  (blue line) and $\kappa_- = 0$ (green line). Note that, depending on the wavevector of the surface wave, it could be observed both in the ``evanescent'' and ``ghost'' portions of the phase space (see panel (c)).

\section{Two classes of Dyakonov surface waves}

\begin{figure}[hbt]
\includegraphics[width=3.5 in]{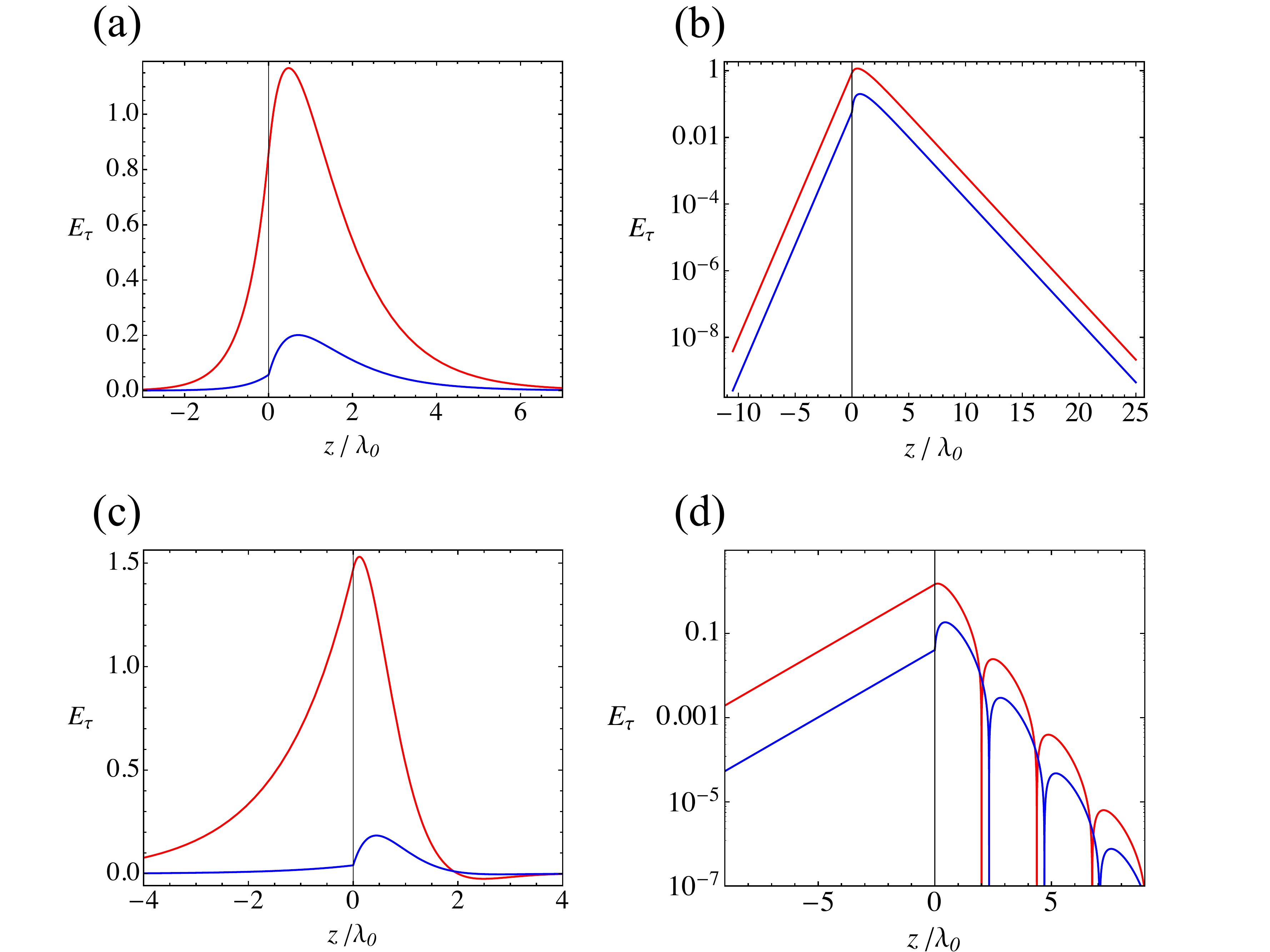}
\caption{\label{fig:SWclasses}
Dyakonov surface waves at the arsenic trisulfide ($n_x = 2.4$, $n_y = 3.02$, $n_z = 2.81$) -- aluminum arsenide ($n_0 = 2.87$) interface : ``conventional'' (a,b) vs. ``ghost'' (c,d), in linear (panels (a),(c)) and logarithmic (panels (b),(d)) scale. Red and blue curves correspond to the projections of the in-plane electric field ${\bf E}_\tau = (E_x, E_y)$ onto the parallel (blue) and perpendicular to (red) to the in-plane momentum ${\bf q}$ directions. The in-plane propagation angle $\theta$ is equal to $24^\circ$ (panels (a),(b)) and $26^\circ$ (panels (c),(d)). The biaxial arsenic trisulfide is on the right of the interface $z = 0$, and the isotropic aluminum arsenide fills the half-space $z < 0$. Note the contrast of the simple exponential decay of the conventional Dyakonov waves in the biaxial medium (see panel (b)) with the oscillatory behavior of the ghost surface waves (panel (d)).
}
\end{figure}

Near the boundary of an isotropic medium with a uniaxial dielectric, the Dyakonov surface wave is formed by evanescent waves on both sides of the interface. However, for a biaxial dielectric that supports both the evanescent and the ghost waves (see Section II), the localized surface wave can be formed from either the evanescent or from ghost waves, depending on its in-plane momentum. As a result, for the interface of a isotropic medium with a biaxial medium, we can have two different types of the Dyakonov surface wave. A ``conventional'' Dyakonov surface, as originally described by M. Dyakonov in 1988 \cite{Dyakonov1}  wave monotonically decays on both sides of the interface, while the ghost surface wave, together with the exponential decay also shows oscillatory behavior in the anisotropic medium -- see Fig. \ref{fig:SWclasses}.

Note that, depending on the magnitude of the permittivity of the isotropic medium $\epsilon_0$ ($\epsilon_z < \epsilon_0  < \epsilon_y$), at a single frequency the isotropic - biaxial interface can either support both the ``conventional'' and the ``ghosts'' mode patterns, or only the "conventional" modes. The corresponding critical value $\epsilon_c$  of the permittivity $\epsilon_0$ is given by the equation (see Appendix D)
\begin{eqnarray}
& & \left(\epsilon_c - \epsilon_z\right)^2 \left(5 \epsilon_z - 3\left(\epsilon_x + \epsilon_y\right) + \frac{\epsilon_x \epsilon_y}{\epsilon_z} \right) \nonumber \\
& + & \left(\epsilon_c -\epsilon_z\right) \left(\epsilon_x^2 + 4 \epsilon_x \epsilon_y + \epsilon_y^2 - 8 \epsilon_z \left(\epsilon_x + \epsilon_y \right) + 10 \epsilon_z^2 \right) 
\nonumber \\
& + & 3 \epsilon_z \left( \epsilon_y - \epsilon_z\right) \left(\epsilon_z - \epsilon_x\right) \nonumber \\
& + & \frac{2}{\epsilon_z} \left( \left(\epsilon_c - \epsilon_z\right)^2 + \epsilon_z 
\left(2 \left(\epsilon_x + \epsilon_y\right) - 5 \epsilon_z \right) \right) \nonumber \\
& \times & \sqrt{\epsilon_z \left( \epsilon_y - \epsilon_z \right) \left( \epsilon_z - \epsilon_x \right) \left( \epsilon_c - \epsilon_z \right) } = 0,
\label{eq:ec}
\end{eqnarray}
which for $\epsilon_x < \epsilon_z < \epsilon_y$ always has a single solution in the interval $\epsilon_z < \epsilon_c < \epsilon_y$. 

\begin{figure}[hbt]
\includegraphics[width=2.5 in]{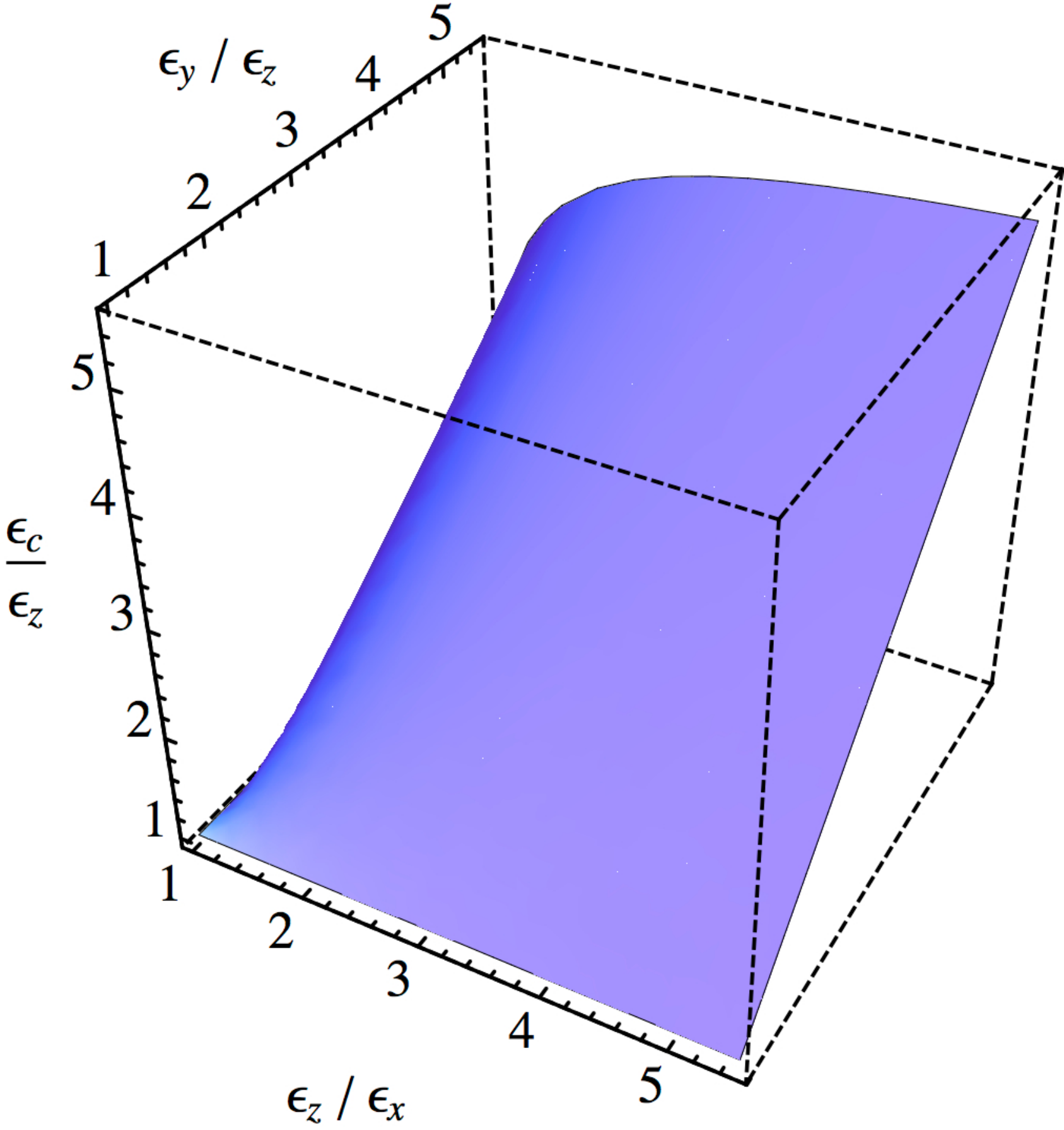}
\caption{\label{fig:ec}
The critical value $\epsilon_c$ of the isotropic medium, in units of $\epsilon_z$, vs.  $\epsilon_z/\epsilon_x$ $\epsilon_y/\epsilon_z$.The dielectric permittivity components of the biaxial medium satisfy $\epsilon_x < \epsilon_z < \epsilon_y$.} 
\end{figure}

In scaled variables $\epsilon_c / \epsilon_z$, $\epsilon_z/\epsilon_x$, $\epsilon_y/\epsilon_z$ the solution of Eqn. (\ref{eq:ec}) can be expressed as
\begin{eqnarray}
\frac{\epsilon_c}{\epsilon_z} & = & {\cal G}\left(\frac{\epsilon_z}{\epsilon_x}, \frac{\epsilon_y}{\epsilon_z}\right). 
\end{eqnarray}
We plot this function in Fig. \ref{fig:ec}.

For $ \epsilon_c < \epsilon_0 < \epsilon_y$, the Dyakonov surface waves that are supported by the interface of isotropic and biaxial dielectric media, belong to the ``conventional'' class for all allowed propagation angles. However, if $\epsilon_z < \epsilon_0 <  \epsilon_c$, for the propagation angle $\theta$ in the range $\theta_1 < \left| \theta \right| < \theta_3$ and $\pi - \theta_3  < \left| \theta \right| < \pi - \theta_1$ we find ``conventional'' Dyakonov waves, while for $\theta_3 < \left| \theta \right|  < \theta_2$ and $\pi - \theta_2 < \left| \theta \right| < \pi - \theta_3$  the surface modes belong to the ``ghost'' class -- see Fig. \ref{fig:projection}(c).  Here, the angle $\theta_3$ only depends on the dielectric permittivies of the media forming the interface, and is defined as the solution of the system of equations (\ref{eq:SW1}) and (\ref{eq:ghost_boundaries}), where the latter taken with the positive signs.

\section{Discussion}

The key feature of the Dyakonov surface waves that makes them an ideal platform for experiments on nonlinear optics and strong coupling, is their inherent ``lossless`` nature. While the residual linear absorption in the dielectric as well as light scattering due to surface roughness can never be completely avoided, the corresponding contributions to the effective mode loss can be dramatically reduced, as demonstrated in Mie resonance experiments with the measured $Q$-factors on the order of $10^{10}$. \cite{Ilchenko} 

As a result, with an evanescent coupling (from e.g. a high-index prism) to the isotropic-biaxial interface, one can observe an enormous increase of the field intensity at this boundary, only limited by the effective loss due to system imperfections (surface and builk disorder, etc.) and ultimately by the non-locality of the dielectric response \cite{LL_media}  (corresponding to the variations of the dielectric permittivity  on the order of $(a_0 / \lambda)^2 \sim 10^{-6}$, where $a_0$ is on the order of the atomic size and $\lambda$ is the wavelength).

For the applications to nonlinear optics however, the effective ``selection rules'' such as the phase-matching conditions \cite{NLO_book} are defined by the spatial variation of the corresponding optical modes. The qualitative difference between the  ``ghost'' and the `conventional'' surface waves, respectively with- and without oscillations away from the interface, that can be simultanenouls supported by the same isotropic-biaxial interface at the same frequency, will therefore have dramatic effect on the nonlinear-optical phenomena in this system. \cite{EN_ghosts}

\section{Conclusions}

In summary, we have developed a complete analytical theory of Dyakonov surface waves at the interface of an isotropic medium with a biaxial anisotropic dielectric. As opposed to earlier work on this subject, our approach
does not require any numerical root-finding, and offers substantial advantage in the description of the surface waves near the propagation threshold. We have also presented a detailed description of the ghost waves that combine the properties of propagating and evanescent solutions, and of the corresponding surface modes supported by these ghost waves.

\section{Acknowledgements}

This work was partially supported by the Army Research Office (grant W911NF-14-1-0639), National Science Foundation (grants DMR-1120923 and DMR-1629276), and  Gordon and Betty Moore Foundation.

\appendix

\section{}

Some of the necessary conditions  for the existence of the Dyakonov wave in (\ref{eq:eps_bounds}) can be immediately obtained from the general structure of Eqn. (\ref{eq:SW1}) and its constituents.
Eqn. (\ref{eq:limit2}) immediately implies that both the first and the last terms in the curly brackets in Eqn. (\ref{eq:SW1}) are negative-definite, therefore
\begin{eqnarray}
\frac{\epsilon_0 + \epsilon_x}{\epsilon_0} \ q_x^2 +
 \frac{\epsilon_0 + \epsilon_y}{\epsilon_0} \ q_y^2 <
 \left(\epsilon_x + \epsilon_y \right) \left(\frac{\omega}{c}\right)^2.
 \label{eq:ineq2}
\end{eqnarray}
Since
\begin{eqnarray}
\frac{\epsilon_x}{\epsilon_0} q_x^2 + \frac{\epsilon_y}{\epsilon_0} q_y^2&  = &  \frac{\epsilon_x \epsilon_y}{\epsilon_0} \left(\frac{q_x^2}{\epsilon_y} + \frac{q_y^2}{\epsilon_x} - \left(\frac{\omega}{c}\right)^2\right) + \frac{\epsilon_x \epsilon_y}{\epsilon_0} \left(\frac{\omega}{c}\right)^2 \nonumber 
\\
& > &  \frac{\epsilon_x \epsilon_y}{\epsilon_0} \left(\frac{\omega}{c}\right)^2,
\label{eq:ineq3}
\end{eqnarray}
from (\ref{eq:limit0}) and (\ref{eq:ineq2}) we obtain
\begin{eqnarray}
\epsilon_x + \epsilon_y > \epsilon_0 + \frac{\epsilon_x \epsilon_y}{\epsilon_0}, 
\label{eq:ineq4}
\end{eqnarray}
which implies that
\begin{eqnarray}
{\rm min}\left(\epsilon_x, \epsilon_y \right) < \epsilon_0 < {\rm max}\left(\epsilon_x, \epsilon_y\right).
\label{eq:bound1}
\end{eqnarray}
Similarly, from (\ref{eq:limit1}), (\ref{eq:ineq2}) and (\ref{eq:ineq3}) 
\begin{eqnarray}
\epsilon_x + \epsilon_y > \epsilon_z + \frac{\epsilon_x \epsilon_y}{\epsilon_0},
\label{eq:ineq5}
\end{eqnarray}
or
\begin{eqnarray}
\epsilon_z & < & {\rm max}\left(\epsilon_x, \epsilon_y\right) - {\rm min}\left(\epsilon_x, \epsilon_y\right) \cdot \left(1 - \frac{{\rm max}\left(\epsilon_x,\epsilon_y\right)}{\epsilon_0}\right) \nonumber \\
& <  & {\rm max}\left(\epsilon_x, \epsilon_y\right).
\label{eq:ineq6}
\end{eqnarray}

\section{}

First, we consider the number of real positive solutions of Eqn. (\ref{eq:poly1}). Since
\begin{eqnarray}
\epsilon_1 + 2 \epsilon_2 & = & \frac{\left(\epsilon_0 - \epsilon_x\right)\left(\epsilon_y - \epsilon_0\right)}{\epsilon_0}  \nonumber \\
& + & \epsilon_x \epsilon_y\left(\frac{1}{\epsilon_z} - \frac{1}{\epsilon_0}\right) > 0,
\end{eqnarray}
since with our choice of $\epsilon_x < \epsilon_y$ (see (\ref{eq:constraintsXY})) the requirement (\ref{eq:eps_bounds}) reduces to 
\begin{eqnarray}
\epsilon_x \leq \epsilon_z < \epsilon_0 < \epsilon_y,
\label{eq:constraints2}
\end{eqnarray} 
and therefore
\begin{eqnarray}
a_4 & > & 0. \label{eq:a4ineq}
\end{eqnarray}
Similarly, since $\hat{u} > 0$,
\begin{eqnarray}
a_1 & < & 0,
\end{eqnarray}
and
\begin{eqnarray}
a_3 & = & \sqrt{\frac{\epsilon_x \epsilon_y}{\epsilon_0} \hat{u}} \cdot 2 \epsilon_0 \left(\frac{\epsilon_x + \epsilon_y - \epsilon_0}{\epsilon_z}   - \frac{\epsilon_x \epsilon_y}{\epsilon_0^2}\right) \nonumber \\
& = & \frac{2}{\epsilon_z} \  \sqrt{\frac{\epsilon_x \epsilon_y}{\epsilon_0} \hat{u}} \cdot  \left(\left(\epsilon_0 - \epsilon_x\right)\left(\epsilon_y - \epsilon_0\right)   + \frac{ \epsilon_x \epsilon_y}{\epsilon_0} \left(\epsilon_0 - \epsilon_z\right) \right) \nonumber \\ &  >&  0. \label{eq:a3ineq}
\end{eqnarray}
Therefore, regardless of the sign of $a_2$,  the number of sign changes of the polynomial $a_4 \chi^4 + a_3 \chi^3 + a_2 \chi^2 + a_1 \chi + a0$ is equal to one if $a_0<0$ and to two if $a_0 > 0$. According to the Descartes' rule of signs, \cite{Descartes} Eqn. (\ref{eq:poly1}) has no more than one positive real  root in the former case and no more than two positive real roots in the latter. So, in general Eqn. (\ref{eq:poly1}) has no more than two positive real roots.

However, the solution of Eqn. (\ref{eq:poly1}) must also satisfy the inequality (\ref{eq:chi_c}). Introducing the new variable
\begin{eqnarray}
\xi \equiv \chi - \sqrt{\epsilon_0 - \epsilon_z},
\end{eqnarray}
to satisfy (\ref{eq:chi_c}) we need $\xi > 0$. From (\ref{eq:poly1}) we obtain
\begin{eqnarray}
a_4 \  \xi^4 + b_3 \  \xi^3 + b_2\  \xi^2 + b_1 \  \xi + b_0 & = & 0,
\label{eq:poly2}
\end{eqnarray}
where
\begin{eqnarray}
b_3 & = & a_3 + 4 a_4 \sqrt{\epsilon_0 -  \epsilon_1}, \\
b_2 & = & a_2+ 3 a_3 \sqrt{\epsilon_0 - \epsilon_z} + 6 a_4 \left(\epsilon_0 - \epsilon_z\right) , \label{eq:b2} \\
b_1 & = & a_1 + 2 a_2 \sqrt{\epsilon_0 - \epsilon_z}  + 3 a_3 \left(\epsilon_0 - \epsilon_z\right) \nonumber \\
& + & 4 a_4 \left(\epsilon_0 - \epsilon_z\right)^{3/2}, \label{eq:b1} \\
b_0 & = & a_0 + a_1 \sqrt{\epsilon_0 - \epsilon_z} + a_2 \left(\epsilon_0 - \epsilon_z\right) \nonumber \\
& + & a_3 \left(\epsilon_0 - \epsilon_z\right)^{3/2} + a_4 \left(\epsilon_0 - \epsilon_z\right)^2.
\end{eqnarray}
From (\ref{eq:a4ineq}) and (\ref{eq:a3ineq})
\begin{eqnarray}
b_3 > 0.
\end{eqnarray}
For $b_0$ we obtain
\begin{eqnarray}
b_0 & = & - \frac{\epsilon_ 0 - \epsilon_z}{\epsilon_0 \epsilon_z} \bigg( \epsilon_x \epsilon_y \hat{u} + \epsilon_0 \sqrt{ \frac{\epsilon_x \epsilon_y}{\epsilon_z} \left(\epsilon_0 - \epsilon_z\right) \hat{u} }  \nonumber \\
& -&   \left(\epsilon_0 - \epsilon_x\right) \cdot \left(\epsilon_y - \epsilon_0\right) \Big)^2 < 0.
\end{eqnarray}
If  either $b_1 >0$, $b_2 > 0$, or $b_1 < 0$, $b_2 < 0$, or $b1< 0$, $b_2 > 0$, then the number of sign changes of the polynomial $a_4 \chi^4 + a_3 \chi^3 + a_2 \chi^2 + a_1 \chi + a0$ is equal to one, and therefore Eqn. (\ref{eq:poly2}) has no more than one real positive root. It is if and only if $b1> 0$, $b_2 <  0$ that Eqn. (\ref{eq:poly2}) can in principle have two positive real roots. 

For $b1> 0$, $b_2 <  0$, from Eqns. (\ref{eq:b2}), (\ref{eq:b1}) we obtain
\begin{eqnarray}
 a_2 + 3 a_3 \sqrt{\epsilon_0 - \epsilon_z}  + 6 a_4 \left(\epsilon_0 - \epsilon_z\right) &  <&  0, \label{eq:in1} \\
 a_1 + 2 a_2 \sqrt{\epsilon_0 - \epsilon_z}  + 3 a_3 \left(\epsilon_0 - \epsilon_z\right) & & \nonumber \\ 
   +   4 a_4 \left(\epsilon_0 - \epsilon_z\right)^{3/2}& > & 0, \label{eq:in2}
\end{eqnarray}
Then 
\begin{eqnarray}
 a_2&   < &  - 3 a_3 \sqrt{\epsilon_0 - \epsilon_z}  -  6 a_4 \left(\epsilon_0 - \epsilon_z\right), \label{eq:prt1} 
\end{eqnarray} 
and
\begin{eqnarray}
 3  a_3  \left(\epsilon_0 - \epsilon_z\right) 
   +   4 a_4 \left(\epsilon_0 - \epsilon_z\right)^{3/2} &  > & \nonumber \\
    - a_1- 
    2 a_2 \sqrt{\epsilon_0 - \epsilon_z} & > & \nonumber \\
    - a_1 + 2  \left(\epsilon_0 - \epsilon_z\right) \cdot \left(3 a_3 + 6 a_4  \sqrt{\epsilon_0 - \epsilon_z} \right), &  & 
\end{eqnarray}
which yields
\begin{eqnarray}
- 3 a_3 - 8 a_4 \sqrt{ \epsilon_0 - \epsilon_z  } &  > & - \frac{a_1}{\epsilon_0 - \epsilon_z}. \label{eq:prt2}
\end{eqnarray}
With $a_1 < 0$, $a_3 >0$ and $a_4 > 0$, and $\epsilon_0 > \epsilon_z$ (see (\ref{eq:eps_bounds})), the left-hand side of (\ref{eq:prt2}) is negative, while the right-hand size is positive. The system of the inequalities (\ref{eq:in1}),(\ref{eq:in2}) is therefore  inconsistent, and the case $b_1 > 0$, $b_2 < 0$  cannot be realized. Therefore, Eqn. (\ref{eq:poly2}) cannot have more than one positive real root, and Eqn. (\ref{eq:poly1}) cannot have more than one real solution with $\chi > \sqrt{\epsilon_0 - \epsilon_z}$.

\section{}

We define $\theta_1$ as the propagation angle that corresponds to the limiting case of the inequality (\ref{eq:limit2}). In terms of our parameter $\hat{u}$ defined by Eqns. (\ref{eq:u}) and (\ref{eq:hatu}), the bound (\ref{eq:limit2}) corresponds to
\begin{eqnarray}
\hat{u}\left(\theta_1\right)  & = & u\left(\theta_1\right) =  0,
\label{eq:uto0}
\end{eqnarray}
which together with (\ref{eq:kpkp2}) implies that
\begin{eqnarray}
\kappa_-\left(\theta_1\right) & = & 0
\end{eqnarray}
Since
\begin{eqnarray}
a_0\left(\hat{u} = 0\right) = a_1\left(\hat{u} = 0\right) = a_3\left(\hat{u} = 0\right) = 0,
\end{eqnarray}
we obtain
\begin{eqnarray}
 {\cal F}\left(\hat{u}=0\right) &  = &  \sqrt{- \frac{a_2\left(\hat{u} = 0\right)}{a_4 \left(\hat{u} = 0\right)}}  \nonumber \\
& = & \sqrt{\epsilon_0 - \epsilon_z + \frac{\epsilon_2^2}{\epsilon_1 + 2 \epsilon_2}}.
\label{eq:Fatu=0}
\end{eqnarray}
Substituting (\ref{eq:uto0}) into (\ref{eq:th}), we obtain
\begin{eqnarray}
\sin^2\theta_1 & = & \frac{\epsilon_x \epsilon_y}{\epsilon_y - \epsilon_x} \left( \frac{\epsilon_a + 2 \epsilon_2}{\epsilon_0 \left(\epsilon_1 + 2 \epsilon_2 \right) + \epsilon_2^2} - \frac{1}{\epsilon_y} \right),
\end{eqnarray}
leading to our definition of $\theta_1$ in Eqn. (\ref{eq:theta1}). 

Since we defined the $x$- and $y$- directions with $\epsilon_y > \epsilon_s$, the inequality (\ref{eq:limit2}) then implies
\begin{eqnarray}
\theta_1 < \left| \theta\right| < \pi/2,
\label{eq:ineq11}
\end{eqnarray}
or
\begin{eqnarray}
\pi/2 < \left| \theta\right| < \pi - \theta_1.
\label{eq:ineq12}
\end{eqnarray}

The angle $\theta_2$ is defined as the propagation direction of the surface wave corresponding to the limiting case of (\ref{eq:AB}) when the latter turns into the exact equality
\begin{eqnarray}
\hat{A}\left(\theta_2\right)  \  \kappa_+\left(\theta_2\right) \kappa_-\left(\theta_2\right) + \hat{B}\left(\theta_2\right) \ & = & 0.
\label{eq:ABeq0}
\end{eqnarray}
Substituting (\ref{eq:ABeq0}) into (\ref{eq:ABrhs}), we find that either
\begin{eqnarray}
\kappa_0\left(\theta_2\right) & = & 0,
\label{eq:th2kp0}
\end{eqnarray}
or
\begin{eqnarray}
- \frac{\epsilon_x \epsilon_y}{\epsilon_0} \hat{u}\left(\theta_2\right) \left(\frac{\omega}{c}\right)^2  & = & \kappa_+\left(\theta_2\right)  \kappa_-\left(\theta_2\right) 
\label{eq:th2u}
\end{eqnarray}
Since for $\theta$ in the range defined by Eqns. (\ref{eq:ineq11}),(\ref{eq:ineq12}) we find $\hat{u} > 0$, and Eqn. (\ref{eq:th2u}) therefore cannot be satisfied -- so that  (\ref{eq:th2kp0}) is the only option. Then, substituting (\ref{eq:th2kp0}) into Eqn. (\ref{eq:SW1}), we obtain
\begin{eqnarray}
 \left(\epsilon_2 -  \frac{\epsilon_x\epsilon_y}{\epsilon_0} \cdot   \hat{u}\left(\theta_2\right) \right)  \cdot \left(\frac{\omega}{c}\right)^2 &    = &     \kappa_+\left(\theta_2\right) \kappa_-\left(\theta_2\right). \ \ \  \ \ \ \ 
\label{eq:th2kpkp}
\end{eqnarray} 
From (\ref{eq:kpkp2}) we obtain
\begin{eqnarray}
\kappa_+\left(\theta_2\right)  \kappa_-\left(\theta_2\right) & = & \left(\frac{\omega}{c}\right)^2 
\sqrt{\frac{\epsilon_x \epsilon_y}{\epsilon_z} \left(\epsilon_0 - \epsilon_z\right) \hat{u}\left(\theta_2\right)}.
\ \ \ \ \  \label{eq:kpkpth2}
\end{eqnarray}
Substituting (\ref{eq:kpkpth2}) into (\ref{eq:th2kpkp}), we find 
\begin{eqnarray}
\hat{u}\left(\theta_2\right) & = & \frac{\epsilon_0 \epsilon_2 }{\epsilon_x \epsilon_y} 
+ \frac{\epsilon_0^2 \left(\epsilon_0 - \epsilon_z\right)}{2 \epsilon_x \epsilon_y \epsilon_z} \nonumber \\
& \times & \left(1 - \sqrt{1 + \frac{4\cdot   \epsilon_z  \left(\epsilon_0 - \epsilon_x \right) \left(\epsilon_y - \epsilon_0 \right)}{\epsilon_0^2\left(\epsilon_0 - \epsilon_z\right)}}\ \right) \ \ \ \ 
\label{eq:uth2}
\end{eqnarray}
From (\ref{eq:hatu}) and (\ref{eq:th2kp0})
\begin{eqnarray}
\hat{u}\left(\theta_2\right) & = & \frac{\epsilon_0}{\epsilon_y} \cos^2\theta_2 +\frac{\epsilon_0}{\epsilon_x}  \sin^2\theta_2 -1.
\label{eq:hatuth2}
\end{eqnarray}
Substituting (\ref{eq:uth2}) into (\ref{eq:hatuth2}) and using (\ref{eq:eps2}), we find
\begin{eqnarray}
\sin^2\theta_2 & = & \frac{\epsilon_y - \epsilon_0}{\epsilon_y - \epsilon_x} 
+ \frac{\epsilon_0}{2 \epsilon_z}
\frac{\epsilon_0 - \epsilon_z}{\epsilon_y - \epsilon_x} \nonumber \\
& \times &  \left(1 - \sqrt{1 + \frac{4 \cdot   \epsilon_z  \left(\epsilon_0 - \epsilon_x \right) \left(\epsilon_y - \epsilon_0 \right)}{\epsilon_0^2 \left(\epsilon_0 - \epsilon_z\right)}}\ \right). \ \ \ \ \  \ \ 
\label{eq:th21}
\end{eqnarray}
To satisfy Eqn. (\ref{eq:AB}), we therefore need
\begin{eqnarray}
0 <  \left| \theta\right| < \theta_2,
\label{eq:ineq21}
\end{eqnarray}
or
\begin{eqnarray}
 \pi - \theta_2  <  \left| \theta\right| < \pi,
 \label{eq:ineq22}
\end{eqnarray}
Together, (\ref{eq:ineq11}), (\ref{eq:ineq12}) and (\ref{eq:ineq21}), (\ref{eq:ineq22}) are equivalent to (\ref{eq:range}).

\section{}

The critical angle $\theta_1$ corresponds to the point where the iso-frequency curve  corresponding to the Dyakonov surface wave in the $(q_x, q_y)$ space terminates at the line  (\ref{eq:boundary_xy}). We can show that the ghost boundary in the first quadrant, 
\begin{eqnarray}
\sqrt{\frac{\epsilon_z - \epsilon_x}{\epsilon_z} }q_x +
\sqrt{\frac{\epsilon_y - \epsilon_z}{\epsilon_z} } q_y & = & 
\sqrt{\epsilon_y - \epsilon_x} \  \frac{\omega}{c},
\label{eq:bnd_ghst}
\end{eqnarray}
can never cross this point. An assumption that such intersection point $(q_x^{(1)},q_y^{(y)})$, that satisfies both (\ref{eq:boundary_xy}) and (\ref{eq:bnd_ghst}),  may exist, leads to the equation
\begin{eqnarray}
& & \left(q_x^{(1)} - \epsilon_y \frac{\omega}{c} \sqrt{\frac{\left( \epsilon_y - \epsilon_z \right)\left(\epsilon_z - \epsilon_x\right)}{\epsilon_z \left(\epsilon_y - \epsilon_x\right)}} \right)^2 \nonumber \\
& + & \left(\frac{\omega}{c}\right)^2 \cdot \frac{\epsilon_y^2}{\epsilon_z} \cdot \frac{\left(\epsilon_z - \epsilon_x\right)^2}{\left(\epsilon_y - \epsilon_x\right)^2} = 0,
\end{eqnarray}
which cannot be satisfied for any $\epsilon_x < \epsilon_z < \epsilon_z$. As a result, in the first quadrant ($q_x >0$, $q_y > 0$) the ghost boundary is either always above or always below the curve of Eqn. (\ref{eq:boundary_xy}). The ellipse of Eqn. (\ref{eq:boundary_xy}) intersects the positive half of the $q_y$-axis at the point of $\sqrt{\epsilon_x}$, while for the ghost boundary (\ref{eq:bnd_ghst}) the corresponding crossing point is at $\sqrt{\epsilon_z \left(\epsilon_y - \epsilon_x\right)/ \left(\epsilon_z - \epsilon_x\right)} > \sqrt{\epsilon_x}$. In the first quadrant of the ${\bf q}$ space the ghost boundary is therefore always above the elliptical curve of Eqn. (\ref{eq:boundary_xy}). As a result, this boundary, and thus the $\theta_1$ ``edge'' of the iso-frequency curve of the Dyakonov surface wave, is {\it always} in the ``conventional'' regime, with the field characterized by the exponential decay on both sides of the interface. 

As a result, for the system to support the ``ghost'' surface waves, the ghost boundary (\ref{eq:bnd_ghst}) must cross the iso-frequency curve of the Dyakonov waves, Eqn. (\ref{eq:SW1}). The onset of the ghost regime then corresponds to the case when the ghost boundary intersects the iso-frequency curve precisely at its end at the angle $\theta_2$. 

As follows from Eqn. (\ref{eq:th2kp0}), the critical angle $\theta_2$ corresponds to the point where the iso-frequency line  of  the Dyakonov surface wave in the $(q_x, q_y)$ space terminates at the circle \begin{eqnarray}
q_x^2 + q_y^2 & = &\epsilon_0  \frac{\omega^2}{c^2}.
\label{eq:circ0}
\end{eqnarray}
For the intersection point $(q_x^{(2)}, q_y^{(2)})$  of (\ref{eq:circ0}) with the ghost boundary  (\ref{eq:bnd_ghst})
in the first quadrant we obtain
\begin{eqnarray}
q_x^{(2)}  & = & \frac{\omega}{c} \frac{\sqrt{\epsilon_z \left(\epsilon_z - \epsilon_x \right)}+\sqrt{\left(\epsilon_y - \epsilon_z \right)\left(\epsilon_0 - \epsilon_z\right)}}{\epsilon_y - \epsilon_x },
\label{eq:qx2} \\
q_y^{(2)}  & = & \frac{\omega}{c} \frac{\sqrt{\epsilon_z \left(\epsilon_y - \epsilon_z \right)}+\sqrt{\left(\epsilon_z - \epsilon_x \right)\left(\epsilon_0 - \epsilon_z\right)}}{\epsilon_y - \epsilon_x }.
\label{eq:qy2}
\end{eqnarray}
Substituting (\ref{eq:qx2}),(\ref{eq:qy2}) into (\ref{eq:SW1}) and using (\ref{eq:kpkp2}), we obtain
\begin{eqnarray}
& & \left(\epsilon_0 - \epsilon_z\right)^2 \left(5 \epsilon_z - 3\left(\epsilon_x + \epsilon_y\right) + \frac{\epsilon_x \epsilon_y}{\epsilon_z} \right) \nonumber \\
& + & \left(\epsilon_0 -\epsilon_z\right) \left(\epsilon_x^2 + 4 \epsilon_x \epsilon_y + \epsilon_y^2 - 8 \epsilon_z \left(\epsilon_x + \epsilon_y \right) + 10 \epsilon_z^2 \right) 
\nonumber \\
& + & 3 \epsilon_z \left( \epsilon_y - \epsilon_z\right) \left(\epsilon_z - \epsilon_x\right) \nonumber \\
& + & \frac{2}{\epsilon_z} \left( \left(\epsilon_c - \epsilon_z\right)^2 + \epsilon_z 
\left(2 \left(\epsilon_x + \epsilon_y\right) - 5 \epsilon_z \right) \right) \nonumber \\
& \times & \sqrt{\epsilon_z \left( \epsilon_y - \epsilon_z \right) \left( \epsilon_z - \epsilon_x \right) \left( \epsilon_0 - \epsilon_z \right) } = 0,
\end{eqnarray}
which defines the values of the dielectric permittivity of the dielectric media corresponding to the onset of ghost surface waves in the system phase space.

\end{document}